\documentclass[aps,prb,a4paper]{revtex4}
\usepackage{graphicx}
\input{m-pictex.tex}
\usepackage{m-ch-en}
\usepackage{amsmath, amssymb}
\usepackage{nicefrac}
\usepackage{color}
\usepackage{psfrag}
\usepackage{textcomp}

\begin{document}

\def\vec#1{\mathbf{#1}}
\def\det#1{\mid #1 \mid}
\def\etal{\emph{et al.}{}}
\def\sfrac#1#2{\nicefrac{#1}{#2}}
\def\degres{\textdegree}

\title{Tracing the Minimum Energy Path on the Free Energy Surface}
\author{Paul Fleurat-Lessard}
\email{Paul.Fleurat-Lessard@ens-lyon.fr}
\affiliation{Laboratoire de Chimie, UMR CNRS 5182, ENS Lyon, 46 allée d'Italie, 69364 Lyon Cedex 07, France.}
\author{Tom Ziegler}
\email{ziegler@ucalgary.ca}
\affiliation{Department of Chemistry, University of Calgary, University Drive 2500\\
 Calgary, Alberta, Canada T2N 1N4}
\begin{abstract}
The free energy profile of a reaction can be estimated in a molecular dynamic (MD)
approach by imposing a mechanical constraint along
a reaction constraint (RC). Many recent studies have
shown that the temperature can greatly influence the path followed by the reactants.
Here, we propose a practical way to construct the minimum energy path directly
on the free energy surface (FES) at a given temperature. First, we follow
the blue-moon ensemble method to derive the expression of the
 free energy gradient for a given RC.
These derivatives are then used to find the actual minimum energy
 reaction path at finite temperature, in a way similar to the Intrinsic Reaction Path of Fukui on the
potential energy surface [J. Phys. Chem. \textbf{74}, 4161 (1970)].
Once the path is know, one can calculate the free energy profile using thermodynamic integration.
We also show that the mass-metric correction
cancels for many types of constraints, making the procedure easy to use. 
Finally, the minimum free energy path at  300K for  the reaction 
\hbox{
\chemical{CCl_2} + 
 \chemical{H_2C,DOUBLE,CH_2}
\chemical{GIVES}
\startchemical[frame=off,height=fit,width=fit]
\chemical[THREE,B2,-SB1,+SB3,Z1][CCl_2]
\stopchemical}
is compared with a path based on a simple 1D reaction coordinate.
A comparison is also given with the reaction path at 0K.
\end{abstract}

\maketitle

\section{Introduction} \label{SecIntro}

Being able to understand, or better to predict, the 
evolution of a complex system is of critical importance in all areas of
chemistry and biology. In turn, this understanding requires the knowledge
 of not only the mechanism at a microscopic level but also of the free 
energy change associated with the reaction under investigation.
In principle, molecular dynamic simulation can give access to the free 
energy profile of chemical processes, and indeed, free energy simulations
have become a key tool in the study of many chemical and biochemical
problems.\cite{AT87}

However, chemical reactions are usually rare events and it would require
a much too long simulation time for a process to occur without
any bias. Thus, finding an efficient
way to accurately compute the free energy difference for a given
reaction is still a very active field of research.\cite{CP02,PS04,YZ04,Jarzynski97a,Jarzynski97b} 
Several methods have been proposed to evaluate the change of 
the free energy along a given path: free energy perturbation,\cite{Zwanzig54}
umbrella sampling,\cite{TV77} thermodynamic integration\cite{SC91} and, more 
recently, the Jarzynski equality.\cite{Jarzynski97a,Jarzynski97b} (See also
ref. \onlinecite{vGunsteren89} for a review.)

Of particular importance is the understanding of the link between constrained
and unconstrained simulation, which was put on firm basis a decade
ago by Carter \etal{} who introduced the Blue Moon  relation.\cite{CCHK89}
During the past few years, this relations has been refined so that many
formula are now at hand to evaluate the derivative of the
free energy along a given reaction coordinate.\cite{SC98,dOB98,DP01}

  The access to analytical gradients on the potential energy surface
 was a very important
step forward in standard quantum chemistry: it became possible
 to find the optimum geometry
of complex systems, optimizing transition states became easier, 
and frequencies 
could be obtained much faster with a higher accuracy.
 On the other hand,
 despite the fact that the
exact equations for the evaluation of gradients of the free energy surface 
have been available for 
many years, their use has mainly been restricted to the evaluation of the free
energy change along a predefined path. Evaluating such a free energy profile is
common in both chemistry and biology. 
However, it seems that the potential of free energy gradients has not been fully appreciated.

In this study, we propose to use the available equations for the gradient
to explore the free energy surface in much the same way as gradients have been used
to explore the potential energy surface.
Special attention will be given to finding the minimum free energy path.

The account of this investigation is organized as follows: In
 the next section (Sec. \ref{SecTheory})
 we first review the main equations employed to evaluate
the derivative of the free energy along a reaction coordinate in a uniform way.
In Section \ref{SecUse}, the scope of these expressions is extended,
and their actual use in a simulation is discussed.
In Section \ref{SecApp}, they are then  used 
to find a minimum energy path on the free energy surface for the addition of dihalide carbene 
to ethylene. Section \ref{SecConcl} offers the concluding remarks. 

\section{Theory} \label{SecTheory}
\subsection{Intrinsic Reaction Path}

We consider a chemical system composed of N atoms of mass $m_i$
 described by 3N cartesian coordinates $x_i$ with  $i=1,\dots,3N$.
We want to construct the minimum energy path connecting the reactants to the products on the
 free energy surface (FES). On the potential energy surface (PES), K. Fukui\cite{Fukui70,Fukui81} 
has defined such a path: the Intrinsic Reaction Path (IRP). On each point of this path, the 
atomic cartesian coordinates satisfy
\begin{equation}
\dots = \frac{m_i dx_i}{\frac{\partial V}{\partial x_i}} =
\frac{m_j dx_j}{\frac{\partial V}{\partial x_j}}  =
\dots \label{eqIRCCart}
\end{equation}

where V is the potential energy and $m_i$ is the mass of the atom with coordinate $x_i$.

     This equation can be simplified by using mass-weighted cartesian coordinates:
\begin{equation}
     x'_i=\sqrt{m_i} x_i \quad i=1,\dots,3N \label{eqDefMW}
\end{equation}

    In its simplified form,  equation (\ref{eqIRCCart}) reads
\begin{equation}
\dots = \frac{dx'_i}{\frac{\partial V}{\partial x'_i}} =
\frac{dx'_j}{\frac{\partial V}{\partial x'_j}}  =
\dots \label{eqIRCMW}
\end{equation}

    Thus, the IRP corresponds to a steepest descent path using mass-weighted coordinates.

In this work, we want to construct a similar path on the FES. Therefore, we have to 
find a path satisfying
\begin{equation}
\dots = \frac{dx'_i}{\frac{\partial A}{\partial x'_i}} =
\frac{dx'_j}{\frac{\partial A}{\partial x'_j}}  =
\dots \label{eqIRCFES}
\end{equation}
where A stands for the Helmholtz free energy. 

    In order to find this path, we have to calculate the gradient of
 the free energy for each point of this path. In practice, this path will be
discretized  by a
set of $k$ points in the configurational space, i.e., by the set of $k$ molecular geometries:
\begin{equation}
 \vec{x'}^{j} =\left\{ {x'}_i^{j}; i=1,\dots, 3N \right\} , j=1,\dots,k
 \label{eqRP0}
 \end{equation}

\subsection{Generalized coordinates}
Even though one can use the 3N cartesian coordinates to describe the system and its evolution,
it is usually easier to employ a set of 3N-6 generalized internal  coordinates, as well as the
overall rotation and translation of the molecule.
 
Further, chemical intuition tells us that most of the time only
 a few degrees of freedom (reaction coordinates) are sufficient
to describe the reaction path. Thus, it appears natural to split the generalized
coordinates into two sub-sets corresponding to the active coordinates, denoted by
$\boldsymbol{\xi} = \left( \xi_1, \ldots, \xi_r \right)$ and the
inactive coordinates, denoted by $\vec{q} = \left( q_1, \ldots, q_n \right)$.
A more quantitative criterion for this separation will be discussed later.
These two sets are associated with two groups of generalized momenta $\vec{p_q}$
    and $\vec{p_\xi}$, and a velocity vector:
 $$ \vec{v_{q\xi}} = \binom{\vec{\dot{q}}}{\boldsymbol{\dot{\xi}}} $$

The IRP will then be constructed in the subset of
 active coordinates $\boldsymbol{\xi}$.

This separation is similar to the adiabatic separation used in
 quantum dynamics between the
slow modes (corresponding to our active set) and the fast modes
(corresponding to our inactive set).  One common point is that the
 inactive coordinates can vary  along the reaction path. In other
words, being inactive does not mean being frozen: an inactive
coordinate is characterized by the fact that it does \emph{not} contribute
to the direction of the minimum free energy path as
the thermal motions along these coordinates are nearly
 harmonic. However,
motion along the inactive coordinates might contribute to the changes
 in the free energy as the curvature of the harmonic potential changes
 along the free energy path. 
 This point will be illustrated in the application
 section.

It is worth discussing here the meaning of a structure on the
potential and on the free energy surfaces. On the potential energy surface,
a structure corresponds to a stationary point: for such a point,
the derivatives of the potential are zero for all coordinates.
On the free energy surface, we can use a similar description: a 
structure corresponds to a point in which the derivatives of
the free energy are zero for all coordinates. By definition of
the inactive coordinate, they do not contribute to a change in the direction
of the minimum free energy path. Therefore, the derivative of the free
energy along an inactive coordinate is zero: $\frac{\partial A}{\partial q_i} =
0$ for all inactive coordinates. As a consequence,
a structure can be define as a point in which the derivatives of the
free
energy are zero for all \emph{active coordinates}. To complete the
geometrical description, we will use the thermal average
of the inactive coordinates during the molecular dynamic simulation.

\subsection{Notations}
 Many expressions have already been proposed to evaluate the derivative of the free
energy in a predefine direction.\cite{SC98,dOB98,dOB00,DP01,DP02,SK03}
 In this section, we will  recall the expressions
that will be used throughout this work.
For the sake of simplicity, the expression will be given in the case where the active
set contains only one coordinate $\xi$.  It is worth noting that once
the reaction path has been constructed, only one coordinate is needed
because the reaction coordinate can be uniquely define as the mass weighted
curvilinear distance from the reactants to the point of interest along
the path.

    We denote by $\vec{J}$ the Jacobian matrix for the
     transformation  from cartesian coordinates to generalized coordinates.
$\vec{x} \rightarrow (\vec{q}, \xi)$:
     $$ \vec{J} = \left(  
\frac{\partial \vec{x}}{\partial \vec{q}} 
     \frac{\partial \vec{x}}{\partial \vec{\xi}} \right)$$

  and by $\det{\vec{J}}$ its determinant (that is the Jacobian).
  We also define $\vec{J'}$: the Jacobian matrix for the
transformation from mass weighted cartesian coordinates to generalized coordinates.

     We introduced the mass matrix $\vec{A}_{q\xi}$ defined by:
$
\vec{A}_{q\xi} = \vec{J}^t \vec{M} \vec{J}
$

     where $\vec{M}$ is a $3N \times 3N$ diagonal matrix containing all
     atomic masses.

The generalized momenta are related to the velocity vector by:
\begin{equation}
 \vec{p_{q\xi}} = \binom{\vec{p_q}}{p_\xi}
= A_{\vec{q}\xi} \binom{\vec{\dot{q}}}{\dot{\xi}}
= \begin{pmatrix}  \vec{A_q}  \vec{\dot{q}} + \vec{B}_{\xi} \dot{\xi} \\
 \vec{B}_{\xi}^t  \vec{\dot{q}} + C_{\xi} \dot{\xi} \\
 \end{pmatrix}  \label{eqpAq}
\end{equation}

   For further convenience, we decompose $\vec{A}_{q\xi}$ and
   its inverse $\vec{A}_{q\xi}^{-1}$ into blocks:
\begin{align}
\vec{A}_{q\xi} & =  \left(
 \begin{array}{cc}
 \vec{A}_{q} & \vec{B_\xi} \\
\vec{B}_{\xi}^t & C_{\xi} \\
\end{array} \right) 
 &
\vec{A}_{q\xi}^{-1} & =  \left(
 \begin{array}{cc}
 \vec{X}_{q} & \vec{Y_{\xi}} \\
\vec{Y}_{\xi}^t & Z_{\xi} \\
\end{array} \right)  \label{eqAinvblock}
\end{align}

Some properties of these matrices are given in Appendix \ref{AppMassMat}.

 Last, let us write explicitly the form
of $Z_{\xi}$ and $\vec{Y}_{\xi}$:
\begin{align}
 Z_{\xi}  & =  \sum_{i=1}^{3N} \frac{1}{m_{i}}
\frac{\partial \xi}{\partial x_{i}}
\frac{\partial \xi}{\partial x_{i}} 
 =  \sum_{i=1}^{3N} 
\frac{\partial \xi}{\partial x'_{i}}
\frac{\partial \xi}{\partial x'_{i}} \\
 \left(\vec{Y}_{\xi}\right)_j  & =  \sum_{i=1}^{3N} \frac{1}{m_{i}}
\frac{\partial q_j}{\partial x_{i}}
\frac{\partial \xi}{\partial x_{i}} 
 =  \sum_{i=1}^{3N} 
\frac{\partial q_j}{\partial x'_{i}}
\frac{\partial \xi}{\partial x'_{i}} 
\end{align}

\subsection{Gradient of the Free Energy}

 The free energy $A(\xi^*)$ is related to the partition function $Q(\xi^*)$
by $A(\xi^*)=- k_B T \ln \left( Q(\xi^*) \right)$. Therefore,  a prerequisite to 
the determination
of $\dfrac{\partial A}{\partial \xi}$ is the evaluation
of $\dfrac{\partial Q}{\partial \xi}$.

The partition functions is defined by:
\begin{equation}
Q(\xi^*) = \int d\vec{q}\int d\vec{p}_{q} dp_{\xi}
\exp\left(-\beta \vec{H} \right)  \label{eq:PartFunc}
\end{equation}
where $\vec{H}$ is the Hamiltonian associated with our system:
\begin{equation}
\vec{H}(\vec{q},\xi,\vec{p}_{q\xi}) = \frac{1}{2}
\vec{p}_{q\xi}^t \vec{A}_{q\xi}^{-1} \vec{p}_{q\xi} +
V(\vec{q},\xi)
\end{equation}

However, in a molecular dynamic simulation, constraining the reaction coordinate
$\xi$ to remain constant and equal to $\xi^*$ implies imposing the additional 
constraint $\dot{\xi} = 0$. Therefore, the ensemble average during the MD simulation
is not the one needed in equation \eqref{eq:PartFunc} because $p_{\xi}$ is \emph{not}
sampled.
When the reaction coordinate $\xi$ is
constrained to a specific value, $\xi=\xi^*$,  the  Hamiltonian associated with
the system becomes:
\begin{equation}
\vec{H}^c_{\xi^*}(\vec{q},\vec{p}_{q}) = \frac{1}{2}
\vec{p}_{q}^t \vec{A}_{q}^{-1} \vec{p}_{q} + V(\vec{q},\xi^*) \label{eqConstrHam}
\end{equation}

 In the following, we will denote by
\begin{equation}
\left< \mathcal{O} \right>_{\xi^*} = \frac{\int d\vec{q}\int
d\vec{p_{q}} \mathcal{O} \exp\left(-\beta \vec{H}^c_{\xi^*}\right)}
{\int d\vec{q}\int d\vec{p_{q}} \exp\left(-\beta \vec{H}^c_{\xi^*}\right)}
\end{equation}

the average of a function $\mathcal{O}$ over the constrained
ensemble. The notation $\left<  \right>_{\xi^*}$ indicates
that the sampling is done along $\vec{p_q}$ and $\vec{q}$ while $\xi$
remains constant and equal to $\xi^*$. 
 The average for an unconstrained
simulation is:
\begin{equation}
\left< \mathcal{O} \right>
 =   \frac{\int d\vec{q}d\xi\int d\vec{p}_{q\xi} \mathcal{O}
\exp\left(-\beta \vec{H}\right)}{\int d\vec{q}d\xi\int d\vec{p}_{q\xi}
\exp\left(-\beta \vec{H}\right)}
\end{equation}

The first step in evaluating the derivative of the free energy is to relate
these two averages.

\subsubsection{ Blue Moon Correction}
Following the work of Carter \etal,\cite{CCHK89} we can rewrite the kinetic
 part of the unconstrained Hamiltonian as
\begin{equation}
\vec{p}_{q\xi}^t \vec{A}_{q\xi}^{-1} \vec{p}_{q\xi}  = 
\vec{p}_{q}^t\vec{A}_{q}^{-1}\vec{p}_{q} 
  + 
\left(p_\xi+ Z_{\xi}^{-1}\vec{Y}_{\xi}^t\vec{p}_{q} \right)^t
 Z_{\xi}\left(p_\xi+ Z_{\xi}^{-1}\vec{Y}_{\xi}^t\vec{p}_{q}
\right) \label{eqKinPart}
\end{equation}

We can then rewrite the 
average value of an operator $\mathcal{O}$ independent of $p_{\xi}$:
\begin{multline}
  \int d\vec{q}\int d\vec{p}_{q\xi} \mathcal{O}
\exp\left(-\beta \vec{H}\right)    
 =  \int d\vec{q}\int d\vec{p}_{q} \mathcal{O} \,
 \exp\left(-\beta \left[ \frac{1}{2} \vec{p}_{q}^t\vec{A}_{q}^{-1}\vec{p}_{q} 
+ V(q,\xi^*)\right] \right)  \\ 
\times \int dp_{\xi}
 \exp\left[-\frac{1}{2}\beta \left(p_\xi+ Z_{\xi}^{-1}\vec{Y}_{\xi}^t\vec{p}_{q} \right)^t
 Z_{\xi}\left(p_\xi+ Z_{\xi}^{-1}\vec{Y}_{\xi}^t\vec{p}_{q}
\right) \right] 
\end{multline}

Using the definition of $\vec{H}^c_{\xi^*}$ and the 
properties of Gaussian integrals (see eq. (\ref{eq:Gint0}) of Appendix \ref{AppGint}), it comes:
\begin{equation}
  \int d\vec{q}\int d\vec{p}_{q\xi} \mathcal{O}
\exp\left(-\beta \vec{H}\right)    
 =  \int d\vec{q}\int d\vec{p}_{q} \mathcal{O} \,
 \exp\left(-\beta \vec{H}^c_{\xi^*} \right) Z_{\xi}^{-\sfrac{1}{2}} \label{eqBMoon}
\end{equation}

 Thus, the relation between the constrained and unconstrained simulations reads
\begin{equation}
\left< \mathcal{O} \right>  = \frac{\int d\vec{q}\int d\vec{p}_{q}
 \exp\left(-\beta \vec{H}^c_{\xi^*}  \right) \mathcal{O} Z_{\xi}^{-\sfrac{1}{2}}}
{\int d\vec{q}\int d\vec{p}_{q}
\exp\left(-\beta \vec{H}^c_{\xi^*} \right)  Z_{\xi}^{-\sfrac{1}{2}}}
 = \frac{\left< \mathcal{O}  Z_{\xi}^{-\sfrac{1}{2}} \right>_{\xi^*}}
{\left<  Z_{\xi}^{-\sfrac{1}{2}} \right>_{\xi^*}} \label{eqBlueMoon}
\end{equation}

which corresponds to the standard Blue Moon correction.\cite{CCHK89}

Using equation (\ref{eqBMoon}), the partition functions can now be written:
\begin{equation}
Q(\xi^*) = \int d\vec{q}\int d\vec{p}_{q}
\exp\left(-\beta \vec{H}^c_{\xi^*} \right)  Z_{\xi}^{-\sfrac{1}{2}}
\end{equation}

which leads to:
\begin{equation}
\left.\frac{\partial Q(\xi)}{\partial \xi}\right)_{\xi^*} = \int d\vec{q}\int d\vec{p}_{q}
e^{-\beta \vec{H}^c_{\xi^*}}  Z_{\xi}^{-\sfrac{1}{2}} 
\left(
 -\beta \frac{\partial  \vec{H}^c_{\xi^*}}{\partial \xi}
 - \frac{1}{2} Z_{\xi}^{-1} \frac{\partial Z_{\xi}}{\partial \xi}
\right) \label{DerPartFunc}
\end{equation}

Starting from this equation, two different procedures have been proposed.
In the first one, this equation is expanded  using the properties of the mass matrix
and of the Gaussian integrals. The second one follows more closely the philosophy
of a MD simulation and tries to relate the derivative of the partition function
to the force $\lambda_{\xi}$ acting on the reaction coordinate $\xi$.
These two procedures will be detailed in the next sections.

\subsubsection{Expanding the Hamiltonian}

From equation \eqref{eqConstrHam}, we have:
\begin{equation}
\frac{\partial \vec{H}^c_{\xi^*}}{\partial \xi} = \frac{1}{2} 
\vec{p}_{q}^t \frac{\partial \vec{A}_{q}^{-1}}{\partial \xi} \vec{p}_{q}
 + \frac{\partial V}{\partial \xi} \label{DerConstrHam}
\end{equation}

Plugging  equations (\ref{eq:Gint2}) of Appendix \ref{AppGint} and (\ref{DerConstrHam})
into equation (\ref{DerPartFunc}), one finds:
\begin{equation}
\left.\frac{\partial Q(\xi)}{\partial \xi}\right)_{\xi^*} = \int d\vec{q}\int d\vec{p}_{q}
e^{-\beta \vec{H}^c_{\xi^*}}  Z_{\xi}^{-\sfrac{1}{2}} 
\left(
- \frac{1}{2} \text{Tr}\left( \vec{A_q} \frac{\partial \vec{A}_q^{-1}}{\partial \xi} \right)
-\beta  \frac{\partial V}{\partial \xi} 
  - \frac{1}{2} Z_{\xi}^{-1} \frac{\partial Z_{\xi}}{\partial \xi}
\right) \label{DerPartFunc2}
\end{equation}

Using
 $ \text{Tr}\left( \vec{A_q} \tfrac{\partial \vec{A}_q^{-1}}{\partial \xi} \right) = 
- \frac{\partial \ln \det{\vec{A}_q}}{\partial \xi}$ and equation (\ref{eqDetAqZ}), one
finally gets:
\def\cte{\xi^*}
 \begin{equation}
\left. \frac{\partial A(\xi)}{\partial \xi}\right)_{\xi^*}  =
\frac{\left<  \det{ Z_\xi}^{-\sfrac{1}{2}} 
  \left(
  \frac{\partial V}{\partial \xi}
  - kT \frac{\partial \ln \det{\vec{J}}}{\partial \xi} \right)
  \right>_{\cte}}{\left<
\det{ Z_\xi}^{-\sfrac{1}{2}} \right>_{\cte}} \label{eqdAxi}
\end{equation}

In the case of multiple constraints, the derivative along
the constraint $\xi_i$ reads:\cite{SC98,dOB00}
 \begin{equation}
\left. \frac{\partial A(\vec{\xi})}{\partial \xi_i}\right)_{\vec{\xi^*}}  =
\frac{\left<  \mid \vec{Z_\xi}\mid ^{-\sfrac{1}{2}} 
  \left(
  \frac{\partial V}{\partial \xi_i}
  - kT \frac{\partial \ln \det{\vec{J}}}{\partial \xi_i} \right)
  \right>_{\cte}}{\left<
\mid \vec{ Z_\xi}\mid^{-\sfrac{1}{2}} \right>_{\cte}} \label{eqdAxiM}
\end{equation}

 Although these formulas are exact, 
 they are not as useful as one might have expected 
because they require the knowledge of the full Jacobian matrix.
 This in turns implies
that the full set of generalized coordinates is known,
 which is something usually 
not desirable for a big molecule.

\subsubsection{Relation with the Lagrange multiplier}

  In cases where constructing the full set of generalized
 coordinates $\vec{q}$ is not
desirable, one has to get rid of
 the explicit dependence of the previous equations on the 
Jacobian. This is what is done in the second approach: instead
of relating the derivative of the constrained Hamiltonian $\vec{H}^c_{\xi^*}$
to the Jacobian, we will relate it to the force acting on the reaction
coordinate during the simulation.
 
Indeed, during a MD simulation, one makes use of the ergodicity principle, and
the averages are actually perform over time and not directly over the phase space.
Then, in order to ensure that the reaction coordinate $\xi$  is constant
and equal to $\xi^*$, a modified Lagrangian is used:
\begin{equation}
\mathcal{L^*} =   \underbrace{\frac{1}{2}
\vec{v_{q\xi}}^t \vec{A}_{q\xi} \vec{v_{q\xi}}
 - V(\vec{q},\xi)}_{\mathcal{L}} + \lambda_{\xi}(\xi-\xi^*) \label{eq:Lc}
\end{equation}

where $\mathcal{L}$ is the Lagrangian of the
unconstrained system and $\vec{v_{q\xi}}$ is the velocity vector.
$\lambda_{\xi}$ is the Lagrange multiplier associated with the reaction coordinate $\xi$.
 Its value is adjusted at each step of the simulation so that $\xi=\xi^*$.
 In practice, the SHAKE algorithm was used for our simulations.\cite{SHAKE}
 In this algorithm, $\lambda_{\xi}$ is adjusted to ensure that $\ddot{\xi} =  0$.

For an unconstrained simulation, the equation of motion of the reaction coordinate is:
\begin{equation}
\frac{d}{dt}\left( \frac{\partial \mathcal{L}}{\partial \dot{\xi}}\right)
 = \frac{\partial \mathcal{L}}{\partial \xi} 
\end{equation} 
which gives:
\begin{equation}
C_{\xi} \ddot{\xi} =  \frac{\partial \mathcal{L}}{\partial \xi}
 - \frac{d}{dt}\left( \vec{B}^t_{\xi} \dot{q} \right) \label{eqXiUC}
 -  \frac{d C_\xi}{dt} \dot{\xi}
\end{equation}

Taking into account that $\dot{\xi}=0$ in a constrained simulation, one has:
\begin{align}
C_{\xi} \ddot{\xi} & =  \frac{\partial \mathcal{L^*}}{\partial \xi}
 - \frac{d}{dt}\left( \vec{B}^t_{\xi} \dot{q} \right)  \\
  & =  \frac{\partial \mathcal{L}}{\partial \xi}
 - \frac{d}{dt}\left( \vec{B}^t_{\xi} \dot{q} \right) + \lambda_{\xi}
\end{align}

Demanding that $\ddot{\xi}=0$ in a constrained simulation, we have:
\begin{equation}
\lambda_{\xi} = -\left\{  \frac{\partial \mathcal{L}}{\partial \xi}
 - \frac{d}{dt}\left( \vec{B}^t_{\xi} \dot{q} \right) \right\} \label{eqLambda}
\end{equation}

Using the definition of the Lagrangian, it comes:
\begin{align}
\lambda_{\xi} & = -\left\{  \frac{1}{2}
\dot{\vec{q}}^t \frac{\partial \vec{A}_{q}}{\partial \xi} \dot{\vec{q}}
 - \frac{\partial V}{\partial \xi}
 - \frac{d}{dt}\left( \vec{B}^t_{\xi} \dot{q} \right) \right\}  \\
& =  \frac{1}{2}
\vec{p_q}^t \frac{\partial \vec{A}^{-1}_{q}}{\partial \xi} \vec{p_q}
 + \frac{\partial V}{\partial \xi}
 + \frac{d}{dt}\left( \vec{B}^t_{\xi} \dot{q} \right) \\ 
& =  \frac{\partial \vec{H}^c_{\xi^*}}{\partial \xi}
 + \frac{d}{dt}\left( \vec{B}^t_{\xi} \dot{q} \right) \label{eqLambda2}
\end{align}

Inserting equation \eqref{eqLambda2} in the expression for the derivative of the partition function gives
\begin{equation}
\frac{\partial Q}{\partial \xi} 
= - \beta 
 \int dp_q dq e^{-\beta H^c_{\xi^*}} Z_{\xi}^{-\sfrac{1}{2}} 
\left(  \lambda_{\xi} - \frac{d}{dt}\left( \vec{B}^t_{\xi} \dot{q} \right)
 + \frac{1}{2\beta} Z_{\xi}^{-1}
 \frac{\partial Z_{\xi}}{\partial \xi} \right)   \label{eqDQ1}
\end{equation}

In order to obtain the final formula we need, we have to integrate analytically
the term $ \frac{d}{dt}\left( \vec{B}^t_{\xi} \dot{q} \right)$. This will be done by following
the work of den Otter \etal\cite{dOB98}
First, we use:
\begin{equation}
 Z_{\xi}^{-\sfrac{1}{2}}  \frac{d}{dt}\left( \vec{B}^t_{\xi} \dot{q} \right)
=   \frac{d}{dt}\left(Z_{\xi}^{-\sfrac{1}{2}} \vec{B}^t_{\xi} \dot{q} \right)
+ \frac{1}{2}  \vec{B}^t_{\xi} \dot{\vec{q}} Z_{\xi}^{-\sfrac{1}{2}} Z_{\xi}^{-1} \frac{d Z_{\xi}}{dt} \label{eqDQ2}
\end{equation}

Using the ergodicity principle, we have:\cite{Vilasi01}
\begin{equation}
 \int dp_q dq e^{-\beta H^c_{\xi^*}} 
 \frac{d}{dt}\left(Z_{\xi}^{-\sfrac{1}{2}} \vec{B}^t_{\xi} \dot{q}
 \right) 
= \lim_{\tau \rightarrow \infty} \frac{1}{\tau} \int_0^\tau dt 
 \frac{d}{dt}\left(Z_{\xi}^{-\sfrac{1}{2}} \vec{B}^t_{\xi} \dot{q} \right)
= 0
\end{equation}

As $\dot{\xi}=0$, we have 
 $\frac{d Z_{\xi}}{dt} = \dot{\vec{q}}^t \vec{\frac{\partial Z_{\xi}}{\partial \vec{q}}}$, and thus:
\begin{equation*}
 \int dp_q dq e^{-\beta H^c_{\xi^*}} \vec{B}^t_{\xi} \dot{q}   \frac{d Z_{\xi}}{dt}
 =   \int dp_q dq e^{-\beta H^c_{\xi^*}} \vec{B}^t_{\xi} \dot{\vec{q}} 
      \dot{\vec{q}}^t \vec{\frac{\partial Z_{\xi}}{\partial \vec{q}}}
\end{equation*}
Using equations \eqref{eq:Gint2} and \eqref{Aq3}, we get:
\begin{align}
 \int dp_q dq e^{-\beta H^c_{\xi^*}} \vec{B}^t_{\xi} \dot{q}   \frac{d Z_{\xi}}{dt}
& =  \int dp_q dq e^{-\beta H^c_{\xi^*}}  \frac{1}{\beta} \vec{B}^t_{\xi}  \vec{A}_q^{-1}
 \vec{\frac{\partial Z_{\xi}}{\partial \vec{q}}}  \notag \\
& =  \int dp_q dq e^{-\beta H^c_{\xi^*}} \left( - \frac{1}{\beta} Z_{\xi}^{-1}\vec{Y}_{\xi}^t 
 \vec{\frac{\partial Z_{\xi}}{\partial \vec{q}}} \right) \label{eqDQ3}
\end{align}

Last, 
\begin{equation*}
\frac{\partial Z_{\xi}}{\partial x_i} =
 \sum_j \frac{\partial Z_{\xi}}{\partial q_j} \frac{\partial q_j}{\partial x_i}
+ \frac{\partial Z_{\xi}}{\partial \xi} \frac{\partial \xi}{\partial x_i}
\end{equation*}
and then 
\begin{equation}
 \vec{Y}_{\xi}^t  \vec{\frac{\partial Z_{\xi}}{\partial \vec{q}}}  = 
 - Z_{\xi} \frac{\partial Z_{\xi}}{\partial \xi} +
  \vec{\frac{\partial \xi}{\partial \vec{x'}}}
 \vec{\frac{\partial Z_{\xi}}{\partial \vec{x'}}} \label{eqDQ4}
\end{equation}

Collecting equations (\ref{eqDQ1}-\ref{eqDQ4}),  we have shown
\begin{equation}
 \int dp_q dq e^{-\beta H^c_{\xi^*}} 
 Z_{\xi}^{-\sfrac{1}{2}}  \frac{d}{dt}\left( \vec{B}^t_{\xi} \dot{q} \right)
=  \int dp_q dq e^{-\beta H^c_{\xi^*}}  Z_{\xi}^{-\sfrac{1}{2}}
 \left( \frac{1}{2\beta} Z_{\xi}^{-1} \frac{\partial Z_{\xi}}{\partial \xi}
 - \frac{1}{2\beta Z_{\xi}^2}  \vec{\frac{\partial \xi}{\partial \vec{x'}}}
 \vec{\frac{\partial Z_{\xi}}{\partial \vec{x'}}}\right) \label{eqDQ5}
\end{equation}

Using this last equation, we recover the general
 formula derived by Sprik \etal,\cite{SC98} den Otter \etal\cite{dOB98} and Darve \etal:\cite{DP01}
\begin{equation}
\frac{\partial A}{\partial \xi} = 
\frac{1}{\left<
Z_\xi^{-\sfrac{1}{2}} \right>_{\cte}}
\left<  Z_\xi^{-\sfrac{1}{2}} 
  \left( \lambda_{\xi} + 
   \frac{1}{2\beta Z_{\xi}^2}
 \left\{ \sum_{i=1}^{i=3N}
 \frac{\partial \xi}{\partial x'_i}  \frac{\partial Z_{\xi}}{\partial x'_i} 
\right\} \right) 
  \right>_{\cte} \label{eqdAxiOB}
\end{equation}

This equation is readily evaluated during a simulation because all the terms depend only
on known quantities such as $\xi$ and $Z_{\xi}$.

When many coordinates are constrained, the derivative along
 $\xi_k$ reads:\cite{dOB00,DP02}
\begin{equation}
\frac{\partial A}{\partial \xi_k} = 
\frac{1}{\left<
\mid \vec{Z_\xi}\mid^{-\sfrac{1}{2}} \right>_{\cte}}
\left<  \mid \vec{Z_\xi}\mid^{-\sfrac{1}{2}} 
  \left( \lambda_{\xi_k} + 
   \frac{kT}{2 \mid \vec{Z_{\xi}} \mid}
 \sum_{j=1}^{j=r} \left( \vec{Z_\xi}^{-1} \right)_{kj}
 \left\{ \sum_{i=1}^{i=3N}
 \frac{\partial \xi_j}{\partial x'_i} 
 \frac{\partial \mid \vec{Z_{\xi}}\mid }{\partial x'_i} 
\right\} \right) 
  \right>_{\cte} \label{eqdAxiOBM}
\end{equation}

\subsubsection{Comparing the two procedures}

At first glance equations \eqref{eqdAxi} and  \eqref{eqdAxiOB} seem very different and it is not
obvious that they both refer to the same quantity.
 The connection between these formulas
 is more clearly seen if one rewrites their numerator:
\begin{align*}
\frac{\partial A}{\partial \xi} &  \propto
\left<   Z_\xi^{-\sfrac{1}{2}}
  \left( \left[ \frac{1}{2} 
\vec{p_q}^t \frac{\partial \vec{A}_{q}^{-1}}{\partial \xi} \vec{p_q}
 +  \frac{\partial V}{\partial \xi} \right]
  + \frac{1}{2\beta} Z_{\xi}^{-1}\frac{\partial Z_{\xi}}{\partial \xi}  \right)
  \right>_{\cte} \tag{\ref{eqdAxi}'}\\
\frac{\partial A}{\partial \xi} & \propto
\left<  Z_\xi^{-\sfrac{1}{2}} 
  \left( \left[ \frac{1}{2}
\vec{p_q}^t \frac{\partial \vec{A}_{q}^{-1}}{\partial \xi} \vec{p_q}
 + \frac{\partial V}{\partial \xi} \right]
 + \frac{d}{dt}\left( \vec{B}^t_{\xi} \dot{q} \right) 
 +  \frac{1}{2\beta Z_{\xi}^2}
 \left\{ \sum_{i=1}^{i=3N}
 \frac{\partial \xi}{\partial x'_i}  \frac{\partial Z_{\xi}}{\partial x'_i} 
\right\} \right) 
  \right>_{\cte} \tag{\ref{eqdAxiOB}'}
\end{align*}

Then, inserting  equation \eqref{eqDQ5} into
 (\ref{eqdAxiOB}') leads to (\ref{eqdAxi}').

\section{Practical considerations} \label{SecUse}

The exact formulas for evaluating the free energy derivatives have
been recalled in the previous section. However, actually using them in
a molecular dynamic simulation requires to tackle two problems. The
first one is related to the chemical system under study: one has to 
choose the coordinates that belong to the active space, so that all
the relevant coordinates are considered. 
 The second problem is more
technical: the previous formulas look quite complicated to evaluate
and one might wonder how to compute them efficiently. We will first
propose a way to decide whether a coordinate should be included into
the active space. Then, we will show that in many cases, the previous
formulas can be greatly simplified.

\subsection{Monitoring the active space} 

The first step of a molecular dynamic simulation aiming at calculating
 the
change in the free energy along a minimum energy path is to 
divide the degrees of freedom into active  and 
inactive coordinates. Of course, it is possible to consider
all coordinates active and to calculate the complete set of
derivatives  using the previous
formulas. However, that would require to launch essentially one MD simulation
for each degree of freedom: such a procedure would be quite expensive. 
More, in contrast to the reaction path on the potential energy
surface, not all degrees of freedom are required: many degrees of
freedom will move in a nearly harmonic well. As a consequence, the derivative
of the free energy along these modes is small, and one can safely
discard them.
 Such a case is observed  in the forthcoming application
for the \chemical{CH} distances of the ethylene molecule: the \chemical{CH}
bond length slightly increases as the cyclopropane is formed, but at
 each point
of the reaction path, they move
 in a quasi-harmonic well. Hence, their contribution to the direction
 of the minimum free energy path  can be neglected.

Therefore, one
must select only a restricted set of coordinates. However,
for complex systems undergoing a reaction, chemical intuition might not be sufficient.
We propose here a way to construct the active and the inactive sets,
and to monitor this separation along the construction of the path.
The value of the derivative of the free energy will be
used as a \emph{quantitative} criterion to discriminate between
 active and non-active coordinates. Ideally, this derivative should be zero for an inactive coordinate.
However in practice, the actual reaction coordinate has non zero component
on all coordinates, including inactive coordinates. The criterion will then
be to compare the derivative of the free energy along an inactive coordinate
to a given threshold.  If it is bigger than this threshold 
then one must consider incorporating $q_n$
into the active set.  More, we will show that one can use the data of a simulation
to estimate the derivative of the free energy along an unconstrained coordinate.

Let us consider a system for which the beginning of the minimum free energy path has already 
been constructed as a set of k points
 $\left( \boldsymbol{\xi^0},\ldots, \boldsymbol{\xi^k} \right)$.  The
purpose of this section if to detail the procedure to find the next
point, $k+1$, of the path.  By definition of the
reaction path, this point can be obtained by following the gradient of
the free energy   along a small distance $ds$:
$$ \boldsymbol{\xi^{k+1}} = \boldsymbol{\xi^{k}} + \frac{\vec{grad}
  A}{\mid \vec{grad}   A \mid} ds $$
The previous formulas can be applied to the data of the MD simulation
  conducted at point $\boldsymbol{\xi^{k}}$ to calculate the
  derivatives 
of the free energy along the active coordinates:
  $\left\{ \frac{\partial A}{\partial \xi_i} \right\}_{i=1,\ldots,r}$.
Before constructing the point $k+1$, we should update the 
 active and inactive sets: if the derivative of the free energy
  along an active coordinate is zero, then this coordinate should be taken out
  of the active set. The reciprocal question is then: shall we include any of the inactive $q_i$ variable
into the active set in order to better describe the rest of the path ? 
  Let us consider the coordinate $q_n$ as an example.
We will now  give the expression of the derivative of the free energy along this
inactive coordinate $q_n$. The previous formulas cannot be used
directly  because they necessitate that
the coordinate  under study is \emph{constrained} during the
simulation.   Let us denote by $q_n^k$
the particular value of $q_n$ at which we want to calculate the
derivate of the  free energy along
$q_n$: $\frac{\partial A}{\partial q_n}\big)_{q_n^k,\xi}$.
  In order to evaluate this quantity,
 one can follow  the same procedure as before, and finds: 

\begin{align}
\left. \frac{\partial A(q_n,\xi)}{\partial q_n}\right)_{q_n=q_n^k,\xi} &  = 
\frac{\left<  \det{ Z_\xi}^{-\sfrac{1}{2}}  \delta(q_n-q_n^k)
  \left(
  \frac{\partial V}{\partial q_n}
   - kT \frac{\partial \ln \det{\vec{J}}}{\partial q_n}
 \right)
  \right>_{\xi}}{\left<
\det{ Z_\xi}^{-\sfrac{1}{2}} \delta(q_n-q_n^k)\right>_{\xi}} 
\label{eqdAq1}
\end{align}

In this expression, we have explicitly included the delta function
 $\delta(q_n - q_n^k)$ which
ensures that the average corresponds to a conditional sampling of the
 phase space in which  $q_n$ is equal to $q_n^k$. 

If we want to avoid calculating the full set of generalized coordinates, we have to 
derive an expression similar to the equation (\ref{eqdAxiOB}) for the coordinate $q_n$.
An important point here is that, in contrast  to $\xi$, $q_n$ was not constrained during 
the simulation. Therefore, the simulation data contain
the sampling over $p_{q_n}$ that would have been  missing in a 
simulation where  \emph{both} $\xi$ and $q_n$ would have been constrained.

The expression for the derivative of the free energy along this unconstrained 
coordinate $q_n$ will be done in two steps. In the first step, we will establish
 the expression for the derivative of the free energy along an unconstrained reaction coordinate,
that is during a simulation without the constraint $\dot{\xi}  =0$. Then, we will use the Blue
Moon relation to obtain the expression of the free energy along a 
non active, unconstrained, coordinate $q_n$ during a simulation with a constrained
reaction coordinate $\xi$.

First, using equation \eqref{eqpAq}, we have:
\begin{equation}
\begin{pmatrix} \dot{\vec{q}} \\ \dot{\xi} \end{pmatrix}
= \vec{A}_{q\xi}^{-1}
\begin{pmatrix} \vec{p_q} \\ p_{\xi} \end{pmatrix}
= \left\{ 
\begin{matrix} \vec{X_q} \vec{p_q} + \vec{Y}_{\xi} p_{\xi} \\
\vec{Y}_{\xi}^t \vec{p_q} + Z_{\xi} p_{\xi} \\
 \end{matrix} \right.
\end{equation}

Using this equation and eq. \eqref{Aq3}, we can rewrite $p_{\xi}$:
\begin{equation}
p_{\xi}   =   - Z_{\xi}^{-1}   \vec{Y}^t_{\xi}\vec{p_q}  + Z_{\xi}^{-1} \dot{\xi} 
 = \vec{B}_{\xi}^t \vec{A}_q^{-1} \vec{p_q}  + Z_{\xi}^{-1} \dot{\xi} 
\end{equation}

Remembering $\dfrac{\partial \vec{H}}{\partial \xi} = - \dfrac{d}{dt}\left( p_{\xi} \right)$,
one finds:
\begin{equation}
\frac{\partial Q}{\partial \xi}
 = \beta \left\{ \int dp_q dq  dp_{\xi}  e^{-\beta H} \frac{d}{dt}\left( \vec{B}_{\xi}^t \vec{A}_q^{-1} \vec{p_q} \right) 
+  \int dp_q dq  dp_{\xi} e^{-\beta H} \frac{d}{dt}\left(
 Z_{\xi}^{-1} \dot{\xi} \right)
\right\}
\end{equation}

The ergodicity principle allows us to say that the first integral is zero.
Therefore, we have:
\begin{align}
\frac{\partial Q}{\partial \xi}
& = \beta  \int dp_q dq  dp_{\xi} e^{-\beta H} \frac{d}{dt}\left(  Z_{\xi}^{-1} \dot{\xi} \right) \notag \\
& = \beta   \int dp_q dq  dp_{\xi} e^{-\beta H} \left(  Z_{\xi}^{-1}
\ddot{\xi} -  Z_{\xi}^{-2} \dot{\xi} \frac{d Z_{\xi}}{dt}\right) \label{eqDerQDP}
\end{align}

We shall now pursue the derivation following a procedure similar to that of  Darve \emph{et al.}\cite{DP01,DP02} 

The second term of equation  \eqref{eqDerQDP} reads:
\begin{align}
  \int dp_q dq  dp_{\xi} e^{-\beta H}  Z_{\xi}^{-2} \dot{\xi}  \frac{d Z_{\xi}}{dt} 
& =  \int dp_q dq  dp_{\xi} e^{-\beta H}  Z_{\xi}^{-2} \dot{\xi} 
    \vec{\frac{\partial Z_{\xi}}{\partial \vec{x'}}} \vec{p_{x'}} \notag \\
& =  \int dp_q dq  dp_{\xi} e^{-\beta H}  Z_{\xi}^{-2} \dot{\xi} 
   \vec{\frac{\partial Z_{\xi}}{\partial \vec{x'}}} 
\left( \vec{J'}^{-1}_{\xi} p_{\xi} + \vec{J'}^{-1}_q \vec{p_q} \right) \label{eqDerQDP2}
\end{align}

We now introduce a new set of generalized momenta $(\widetilde{\vec{p_q}}, \widetilde{p_{\xi}})$
defined by:
\begin{equation}
\left\{
\begin{aligned}
\widetilde{\vec{p_q}} &=\vec{p_q} \\
\widetilde{p_{\xi}} & = Z_{\xi}^{-1} \dot{\xi} = p_{\xi} + Z_{\xi}^{-1} \vec{Y}_{\xi}^t \vec{p_q}
\end{aligned}
\right.
\end{equation}
This transformation is valid as it is invertible.
 Moreover, the Jacobian $\det{\widetilde{\vec{J}}}$ associated with it is
equal to 1. 
With these new variables, equation \eqref{eqKinPart} reads:
\begin{equation}
 \frac{1}{2} \vec{p_{q\xi}}^t \vec{A}_{q\xi}^{-1} \vec{p_{q\xi}} 
=   \frac{1}{2} \widetilde{\vec{p_{q}}}^t \vec{A}_{q}^{-1} \widetilde{\vec{p_{q}}}
 +   \frac{1}{2} \widetilde{p_{\xi}} Z_{\xi}  \widetilde{p_{\xi}} 
\end{equation}

Equation \eqref{eqDerQDP2} then reads
\begin{multline}
  \int dp_q dq  dp_{\xi} e^{-\beta H}  Z_{\xi}^{-2} \dot{\xi}  \frac{d Z_{\xi}}{dt} 
 =  \int dq e^{-\beta V} 
\int d\widetilde{p_q} e^{-\frac{1}{2} \beta \widetilde{\vec{p_{q}}}^t \vec{A}_{q}^{-1} \widetilde{\vec{p_{q}}}} \\
\times \int d\widetilde{p_{\xi}} \det{\widetilde{\vec{J}}} 
  e^{-\frac{1}{2} \beta \widetilde{p_{\xi}} Z_{\xi}  \widetilde{p_{\xi}}} 
 Z_{\xi}^{-1} \widetilde{p_{\xi}}  \vec{\frac{\partial Z_{\xi}}{\partial \vec{x'}}} 
\left[  \vec{J'}^{-1}_{\xi} \widetilde{p_{\xi}} + 
\left( \vec{J'}^{-1}_q - Z_{\xi}^{-1}\vec{Y}_{\xi}^t \right) \widetilde{\vec{p_q}} \right]
\end{multline}

As the Hamiltonian is even in $\widetilde{p_{\xi}}$, after integration over $\widetilde{p_{\xi}}$,
only even terms remain:
\begin{equation}
  \int dq dp_q  dp_{\xi} e^{-\beta H}  Z_{\xi}^{-2} \dot{\xi}  \frac{d Z_{\xi}}{dt} 
 =  \int dq  d\widetilde{p_q} d\widetilde{p_{\xi}} \det{\widetilde{\vec{J}}} 
  e^{-\beta H} 
 Z_{\xi}^{-1} \widetilde{p_{\xi}}^2  \vec{\frac{\partial Z_{\xi}}{\partial \vec{x'}}} 
 \vec{J'}^{-1}_{\xi} 
\end{equation}

 Making use of eq.  \eqref{eq:Gint2} to integrate over $\widetilde{p_{\xi}}$, it comes:
\begin{align}
  \int dq dp_q  dp_{\xi} e^{-\beta H}  Z_{\xi}^{-2} \dot{\xi}  \frac{d Z_{\xi}}{dt} 
& =  \int dq d\widetilde{q} d\widetilde{p_{\xi}} \det{\widetilde{\vec{J}}} 
  e^{-\beta H} 
 \frac{1}{\beta Z_{\xi}^2}   \vec{\frac{\partial Z_{\xi}}{\partial \vec{x'}}} 
 \vec{J'}^{-1}_{\xi} 
 \\
& =  \int dq dp_q   dp_{\xi} e^{-\beta H}  
 \frac{1}{\beta Z_{\xi}^2}   \vec{\frac{\partial Z_{\xi}}{\partial \vec{x'}}} 
\vec{\frac{\partial \xi}{\partial \vec{x'}}}
\end{align}

Collecting the previous equations leads to:
\begin{equation}
\frac{\partial A}{\partial \xi} =  \frac{1}{Q}
 \int dp_q dq  dp_{\xi} e^{-\beta H}  
\left( -Z_{\xi} \ddot{\xi} +  \frac{1}{\beta Z_{\xi}^2} 
  \vec{\frac{\partial Z_{\xi}}{\partial \vec{x'}}} 
\vec{\frac{\partial \xi}{\partial \vec{x'}}} \right)
\end{equation}

This equation is the same as the one obtained by Darve \emph{et al.}
(eq. (24) of ref. \onlinecite{DP02}), but we
arrived at it with different assumptions.

We now apply this equation to the case
of a simulation where $\xi$ is constrained but $q_n$ is not. 
Applying the Blue Moon relation (see eq. \ref{eqBlueMoon}), one 
finds:\cite{DP01,DP02}
\def\dqn{\delta_{q_n}}
\begin{equation}
\left. \frac{\partial A(q_n,\xi)}{\partial q_n}\right)_{q_n^k,\xi}   = 
\frac{1}{\left<
\det{ Z_\xi}^{-\sfrac{1}{2}} \dqn\right>_{\xi}}
\left<  \det{ Z_\xi}^{-\sfrac{1}{2}}  \dqn
  \left( -Z_{q_n}^{-1} \ddot{q_n} +
   \frac{kT}{Z_{q_n}^2}
 \left\{ \sum_{i=1}^{i=3N}
 \frac{\partial q_n}{\partial x'_i}  \frac{\partial Z_{q_n}}{\partial x'_i} 
\right\} \right) 
  \right>_{\xi} 
\label{eqdAq12}
\end{equation}

where  $\delta_{q_n}$ stands for $\delta(q_n-q_n^k)$, 
and $Z_{q_n}$ is the part of the inverse mass matrix corresponding to $q_n$:
$ Z_{q_n} = \sum_i \frac{1}{m_i} \frac{\partial q_n}{\partial x_i}
 \frac{\partial q_n}{\partial x_i}$. As already noted,\cite{DP02}
 this expression is slightly different from  equation
 \eqref{eqdAxiOB}, despite the fact that they both 
relate 
the derivative of the free energy along $q_n$ to
the force acting on  $q_n$ during the MD simulation. 
The origin of this difference comes from the nature of the sampling used to
estimate the two expressions: 
 equation \eqref{eqdAq12}  corresponds to a conditional average
performed during a  simulation in which $q_n$ was \emph{not} constrained
whereas equation  \eqref{eqdAxiOB} corresponds to 
a simulation in which both $\xi$  \emph{and} $q_n$ were constrained.
It is worth stressing here that both expressions are valid
but correspond to  different contexts. 
More, 
as long as the sampling along $q_n$ is sufficient, evaluating
the derivative of the free energy along $q_n$
during a simulation in which only $\xi$
is constrained 
using the conditional averaging of equation  \eqref{eqdAq12}
will lead to the same numerical result as equation  \eqref{eqdAxiOB}
using a MD simulation with both $\xi$ and $q_n$ constrained.

Practical use of this equation is described in Appendix \ref{AppCalcqn}.

To conclude this section, we detail one possible use of
equations \eqref{eqdAxiOB} and \eqref{eqdAq12} for the construction
of a reaction path on the free energy surface. We suppose as before that
the active set is known and denoted by $\boldsymbol{\xi}$, and that 
the path is partially constructed, up to the point $k$ corresponding
to the value $\boldsymbol{\xi^k} = \left\{\xi^k_i\right\}_{i=1,\ldots,r}$ of the active coordinates. To find the next point
 $\boldsymbol{\xi^{k+1}}$, one should:
\begin{enumerate}
\item Launch a simulation while constraining each active
  coordinates  $\xi_i$ ($i=1,\ldots,r$) to be  constant, and equal to $\xi^k_i$,

\item Use formula  \eqref{eqdAxiOB} (or  \ref{eqdAxi}) to compute the derivative of the free
energy along the active coordinates: $\boldsymbol{\frac{\partial
    A}{\partial \xi}}$,
\item   taken out of the active set the coordinates corresponding to
  zeo derivatives,

\item Then,  use the \emph{same simulation data} and equation \eqref{eqdAq12} to 
evaluate the derivative of the free energy along the inactive coordinates
 $\boldsymbol{\frac{\partial A}{\partial q}}$,  then include into the active set
the coordinates associated to non-zero derivatives, 
\item Recollect \emph{all} derivatives to obtain the full gradient:
$ \vec{grad}A =\left\{\boldsymbol{\frac{\partial A}{\partial \xi}}^t,
\boldsymbol{\frac{\partial A}{\partial q}}^t \right\}^t$.
\item Construct the next point of the path  by following the gradient of the
free energy along a small distance $ds$:
\begin{equation*}
\boldsymbol{\xi^{k+1}} = \boldsymbol{\xi^k} + ds  \frac{\vec{grad}
  A}{\mid \vec{grad}   A \mid} 
\end{equation*}
\end{enumerate}

Such a procedure has been used  to study the addition of \chemical{CCl_2}
to ethylene, and is explained in the section \ref{SecApp}.

\subsection{Special types of constraints}

The preceding equations (\ref{eqdAxi}, \ref{eqdAxiM}, \ref{eqdAq1} and \ref{eqdAq12})
 have been derived
 without considering the actual form of the constraints. Therefore,
 they are general but they look quite difficult to compute efficiently. Despite
 the fact that a general procedure has already been given by Darve \etal,\cite{DP01},
 we would like to show here that  many common constraints lead to considerable simplifications 
of the previous expressions. Four cases are described here in which 
 $Z_{\xi}$ is a constant. In those cases,
the previous equations become:
\begin{align}
\frac{\partial A}{\partial \xi}  & = 
  \left< \lambda_{\xi} 
  \right>_{\xi^*}  \label{eqdAxil} \\
\frac{\partial A}{\partial q_n}  & = 
\frac{\left<  \dqn
 \left( -Z_{q_n} \ddot{q_n} \right)
  \right>_{\xi^*}}{\left<  \dqn \right>_{\xi^*}} \label{eqdAqnl} 
\end{align}
\subsubsection{Bond distance}
If the reaction coordinate is chosen to be the bond distance between
atoms $i$ and $j$, we have:\cite{SK03}
\begin{equation}
\left\{
\begin{aligned}
\xi & = d_{ij} = \sqrt{ (x_i-x_j)^2 + (x_{i+1} - x_{j+1})^2 + (x_{i+2}-x_{j+2})^2} \\
Z_{\xi}& =\left( \frac{1}{m_i} + \frac{1}{m_j} \right)
\end{aligned}
\right.
\end{equation}

Care must be taken when constraining many bond distances. In this case, $\vec{Z_{\vec{\xi}}}$ might
no longer be a constant, and the above simplifications should not be applied blindly.
Let us consider a simulation with two constrained bond distances. Two cases can be met depending
on whether these two bonds share a common atom or not.

We consider first the case where the two bonds do \emph{not} share a common atom: for example, 
bonds between atoms $i$ and $j$ and between $k$ and $l$. 
 The matrix $\vec{Z_{\vec{\xi}}}$ is:
\begin{equation}
\vec{Z_{\vec{\xi}}} = \begin{pmatrix} Z_{d_{ij}} & 0 \\ 0 & Z_{d_{kl}} \\ \end{pmatrix}
\end{equation}
with, $Z_{d_{ij}} = \sfrac{1}{m_i}+\sfrac{1}{m_j}$ and
$Z_{d_{kl}} =\sfrac{1}{m_k}+\sfrac{1}{m_l}$.
As a consequence, $\det{\vec{Z_{\xi}}}$ is equal to $\det{Z_{d_{ij}} } \det{ Z_{d_{kl}} }$ and
is a constant: one can use simplified equations \eqref{eqdAxil} and \eqref{eqdAqnl}.

Let us consider now the case where the constrained bonds share a common atom: for
example bonds between atoms $i$ and $j$ and between atoms $j$ and
$k$. We also denote by $\alpha$ the angle between the three atoms:

\begin{center}
\includegraphics{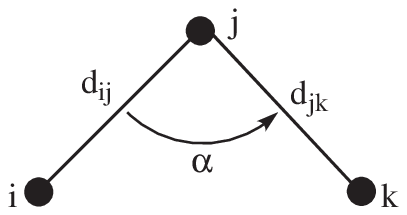}
\end{center}

The matrix $\vec{Z_{\vec{\xi}}}$ reads:
\begin{equation}
\vec{Z_{\vec{\xi}}} = \begin{pmatrix} Z_{d_{ij}} &
 \frac{\cos \alpha}{m_j} \\  \frac{\cos \alpha}{m_j} & Z_{d_{jk}} \\ \end{pmatrix}
\end{equation}
As a consequence,  $\det{\vec{Z_{\xi}}}$  is no longer a constant and equations \eqref{eqdAxiOBM}
and \eqref{eqdAq12} must be used.

Another commonly used reaction coordinate is the difference of two distances. Once again, one has to consider the case
where the two distances share a common atom or not. When the two bonds are not sharing any atom, i.e when they are involving
atoms $i$, $j$, $k$ and $l$, we have:
\begin{equation}
\left\{
\begin{aligned}
\xi & = d_{ij}-d_{kl} \\
Z_{\xi} & =  \frac{1}{m_i} + \frac{1}{m_j} + \frac{1}{m_k} + \frac{1}{m_l} 
\end{aligned}
\right.
\end{equation}

On the other hand, when the two distances share a common atom $j$, the previous equations become:
\begin{equation}
\left\{
\begin{aligned}
\xi & = d_{ij}-d_{jk} \\
Z_{\xi} & =  \frac{1}{m_i} + \frac{1}{m_k} + \frac{2}{m_j}\left( 1 -  \cos \alpha  \right)
\end{aligned}
\right.
\end{equation}
So that $Z_{\xi}$ is not a constant and the correction terms should be evaluated explicitly.

\subsubsection{Generalized distance}
Another quite common reaction coordinate is the mass-weighted distance between
one  reference geometry and the current geometry of the system.\cite{SSM01,SK03}
We note $\left\{y_i; i=1, \ldots, 3N\right\}$ the cartesian coordinates of the reference geometry. We have:
\begin{equation}
\left\{
\begin{aligned}
\xi & = \sqrt{ \sum_{i} m_i (x_i -y_i)^2  } \\
Z_{\xi} & = 1
\end{aligned}
\right.
\end{equation}

 \subsubsection{Bond angle}
  We now consider  constraining
 the angle $\alpha$ between
the atoms $i$, $j$ and $k$, that is the angle between the bonds $ij$ and $jk$.
 We denote by $d_{ij}$, $d_{jk}$ and $d_{ik}$ the distances
between these atoms. With these notations, $Z_{\xi}$ Reads:\cite{WDC55}
\begin{equation}
 Z_{\xi} = Z_{\alpha} = \frac{1}{m_i d_{ij}^2} +  \frac{1}{m_k d_{jk}^2}  +  \frac{d_{ik}^2}{m_j d_{ij}^2 d_{jk}^2}
\end{equation}

Therefore, $Z_{\xi}$ is not a constant and one must use the complete formulas.
However, if the bond distances $d_{ij}$ and $d_{jk}$ are also constrained, the formulas
can be simplified. When considering a simulation with $\alpha$, $d_{ij}$ and
$d_{jk}$ all constrained, $\vec{Z_{\xi}}$ reads\cite{WDC55}
\begin{equation}
\vec{Z_{\xi}} = \begin{pmatrix}
Z_{\alpha} & -\frac{\sin \alpha}{m_j d_{jk}} &  -\frac{\sin \alpha}{m_j d_{ij}} \\
 -\frac{\sin \alpha}{m_j d_{jk}} & Z_{d_{ij}} &  \frac{\cos \alpha}{m_j} \\
 -\frac{\sin \alpha}{m_j d_{ij}} & \frac{\cos \alpha}{m_j} & Z_{d_{jk}} \\
\end{pmatrix} \label{eqZddangle}
\end{equation}

Developing $\det{\vec{Z_{\xi}}}$, one finds: 
\begin{equation}
\det{\vec{Z_{\xi}}} = Z_{\alpha} \left( \frac{1}{m_i m_k} + \frac{1}{m_i m_j} + \frac{1}{m_j m_k} \right)
\end{equation}
Thus, the  derivatives $\frac{\partial}{\partial x_i} \det{\vec{Z_{\xi}}}$ of $\det{\vec{Z_{\xi}}}$ 
 are \emph{not} zero
but $\det{\vec{Z_{\xi}}}$ is constant during a simulation. Using this fact, equation \eqref{eqdAxiOBM}
reads:
\begin{align}
\frac{\partial A}{\partial \xi_k} &= 
\frac{1}{Q^c_{\cte}} 
\left( 
\left<   \lambda_{\xi_k} \right> + 
   \frac{k \left<T\right>}{2 Z_{\alpha} } B_k \right) \label{eqZdist}\\
\text{with } B_k &=
 \sum_{j=1}^{j=r} \left( \vec{Z_\xi}^{-1} \right)_{kj}
 \left\{ \sum_{i=1}^{i=3N}
 \frac{\partial \xi_k}{\partial x'_i} 
 \frac{\partial Z_{\alpha} }{\partial x'_i} 
\right\} 
\end{align}

As  $\alpha$, $d_{ij}$ and $d_{jk}$ are all constrained, $Z_{\alpha}$ and thus $B_k$
are easily computed during (or after) the simulation.
We conclude the discussion of this case by noting that despite the fact that one
should take the corrective terms $ \frac{k \left<T\right>}{2 Z_{\alpha} } B_k$ into 
account, they are of the order of magnitude of some tenth of $kT$ and thus
one might wonder if they are negligible or not. This point will be discussed in greater details in the next section.

\subsubsection{Linear Constraint}
 Linear constraint can be written as
$\xi=\sum_{i}a_{i}\vec{x_{i}}$ and thus, $Z_{\xi} = \sum_{i} m_i^{-1} a^2_{i}$ is a constant.

This type of constraint has already been used in order to
calculate the free energy change along the instrinsic reaction path
constructed on the PES.\cite{MZ01}

These constraints are of considerable practical importance. First,
 in a simulation,
  it is quite common to constrain also the global rotation and the
 global translation of the molecule. 
  Equations similar to the previous equations \eqref{eqdAxi} and \eqref{eqdAq1} can be
  derived by replacing $\det{Z_{\xi}}$ by
 $\det{\vec{Z_{\xi,\vec{T},\vec{R}}}}$. By construction,
the global rotation and translation are orthogonal to all internal
coordinates. Therefore, we have
 $=\det{Z_{\xi}}\det{\vec{Z}_{\vec{T},\vec{R}}}$. 

We show in appendix \ref{AppTR} that constraining the overall rotation and translation
corresponds to applying six additional \emph{linear constraints} on the system.

 As a consequence, the term $\det{\vec{Z}_{\vec{T},\vec{R}}}$ is a constant that can be ignored when 
 calculating the derivatives of the free energy, leading to the following formulas:
\begin{align}
\left.\frac{\partial A}{\partial \xi}\right)_{\xi^*}  & = 
\frac{\left< 
 Z_{\xi}^{-\sfrac{1}{2}}   \left( 
  \frac{\partial V}{\partial \xi}
  - kT \frac{\partial \ln \det{\vec{J}}}{\partial \xi} \right)
  \right>_{\xi^*,T,R}}{\left<  Z_{\xi}^{-\sfrac{1}{2}}
 \right>_{\xi^*,T,R}}\label{eqdAxilTR} \\
\frac{\partial A}{\partial \xi} & = 
\frac{1}{\left<
Z_\xi^{-\sfrac{1}{2}} \right>_{\cte}}
\left<  Z_\xi^{-\sfrac{1}{2}} 
  \left( \lambda_{\xi} + 
   \frac{1}{2\beta Z_{\xi}^2}
 \left\{ \sum_{i=1}^{i=3N}
 \frac{\partial \xi}{\partial x'_i}  \frac{\partial Z_{\xi}}{\partial x'_i} 
\right\} \right) 
  \right>_{\xi^*,T,R} \label{eqdAxiOBTR}
\end{align}
Second, let us consider
 a simulation with many active coordinates, \emph{all} described by linear constraints 
(which may include the translation and rotation constraints).
Let us denote by $\left\{\xi_i; i=1,\ldots,r \right\}$ the r
 linear constraints applied during the simulation. We have
\begin{align}
\xi_{i} &= \sum_{j=1}^{j=3N} c_{ij} x_{j} & \text{For i in} 1, \ldots, r
\end{align}
The Lagrangian associated with this simulation is:
\begin{equation}
\mathcal{L^*} = \frac{1}{2} \vec{\dot{x}}^t \vec{M} \vec{\dot{x}} - V(\vec{x})
 + \sum_{i=1}^{r} \lambda_i \left( \xi_i - {\xi_i}^* \right) 
\end{equation}
The equations of motion are then:
\begin{align}
 m_{j}\ddot{x_{j}} & =  - \frac{\partial V}{
  \partial x_{j}} + \sum_{i=1}^{i=r} \lambda_{i} \frac{\partial
  \xi_{i}}{\partial x_{j}} 
 &  \text{For j in } 1, \ldots, 3N
  \end{align}
  Demanding that $\ddot{\xi_i}=0$ for all $i$ in $1, \ldots,r$ leads to a set of coupled linear equations:
  \begin{align}
   \sum_{j=1}^{3N} - \frac{1}{m_{j}} \frac{\partial V}{
  \partial x_{j}} c_{ij}+ \sum_{k=1}^{k=r} \lambda_{k} \sum_{j=1}^{j=3N} \frac{c_{kj}c_{ij}}{m_{j}} & = 0
  & \text{For i in } 1, \ldots, r \label{eqSys}
  \end{align}
  To find the forces acting on the constraints, i.e. to find the values of all $\lambda_k$,
  we have to solve this linear system. We define:
  \begin{align} 
  D_{i}  & =  \sum_{j} \frac{1}{m_{j}} \frac{\partial V}{
  \partial x_{j}} c_{ij} & \text{For i in } 1, \ldots, r  \label{eqLbeg} %
  \end{align}
 Using the definition of $\vec{Z_\xi}$, it is easily seen that the term $\sum_{j=1}^{j=3N} \frac{c_{kj}c_{ij}}{m_{j}}$
 corresponds to the element $\left[\vec{Z_{\xi}}\right]_{ik}$.
  In matrix notation, equation \eqref{eqSys} reads
  \begin{equation}
  \vec{Z_{\xi}} \vec{\lambda} = \vec{D} \label{eqLend}
  \end{equation}

  which is easily solved by inverting the matrix $\vec{Z_{\xi}}$ which
 depends only on the  definition of the linear constraints $\xi_{i}$.
 This last expression can be further simplified when the constraints
 are expressed in the mass-weighted cartesian coordinates frame: 
 \begin{align}
 \xi_i &= \sum_{j=1}^{j=3N} \frac{c_{ij}}{\sqrt{m_j}} x'_j  & \text{For i in } 1, \ldots, r
 \end{align}
 Requesting that the linear constraints form an orthonormal set in the mass weighted basis leads to:
 \begin{equation}
 \left< \xi_i \mid \xi_k \right> = \sum_{j=1}^{j=3N} \frac{c_{ij} c_{kj}}{m_j} = \delta_{ik}
 \end{equation}
    Therefore, in the case of orthogonal constraints, the inverse of the mass matrix $Z_\xi$ reduces
    to the identity matrix and 
      the previous equations (\ref{eqLbeg}-\ref{eqLend}) become
  \begin{equation}
  \lambda_{i} = \sum_{j=1}^{3N} \frac{1}{m_{j}} \frac{\partial V}{
  \partial x_{j}} c_{ij} \label{eqL2}
  \end{equation}
More, equation \eqref{eqdAxiOBM}, that should be used to calculate the
free energy derivatives in the case of multiple active coordinates, 
reduces to its simplest form,
similar to \eqref{eqdAxil}:
  \begin{align}
\frac{\partial A}{\partial \xi_k}  & = 
  \left< \lambda_{\xi_k} 
  \right>_{\boldsymbol{\xi^*}} &\text{For k in } 1, \ldots, r
\end{align}
This illustrates one convenient property of orthogonal linear
constraints: they are all decoupled which leads to considerable
simplification for the calculation of the free energy derivatives.

However, the most interesting aspect of these constraints 
appears when one consider that the \emph{full} set of generalized coordinates
consists of linear coordinates:
\begin{align}
\xi_i &= \sum_{j=1}^{j=3N} \frac{c_{ij}}{\sqrt{m_j}} x'_j & \text{For i in } 1, \ldots, r \\
q_k &= \sum_{j=1}^{j=3N} \frac{c_{kj}}{\sqrt{m_j}} x'_j & \text{For k in } r+1, \ldots, 3N 
\end{align}
Using the previous equations, one finds that the force acting on an inactive
coordinate $q_k$ during the MD simulation is 
  \begin{align}
   F_{q_k} &= \sum_{j=1}^{j=3N} \frac{1}{m_{j}} \frac{\partial V}{
  \partial x_{j}} c_{kj} &\text{For k in } r+1, \ldots, 3N \label{eqL3}
  \end{align}

Comparing equations \eqref{eqL2} and \eqref{eqL3} shows that the effect of the Lagrange
multiplier $\lambda_i$ is to exactly compensate the force acting on the active coordinate
$\xi_i$ during the MD simulation, thus ensuring that it remains constant.

These last equations show the  advantage of   linear constraints
over other constraints:   one can estimate  the gradient of the free energy for the \emph{full}
  set of generalized coordinates by analyzing the
data of \emph{one} MD simulation. 
The only condition is to have sufficient sampling   along non active
 coordinates.  Therefore, one can launch a 
simulation with a small active set, ideally comprising only the reaction coordinate, without
loosing any information.

  \section{Application}  \label{SecApp}
   It is worth noticing that once we have an expression from which to calculate the
  free energy derivatives, we can apply the same algorithms as those used
  in quantum chemistry in connection with potential energy gradients.
 For example, it is possible to optimize a structure 
  directly on the free energy surface in the subset of the active
  coordinates. It is further  possible to find
   a transition state along a path, to calculate the Hessian by finite difference, and
  thus to characterize the structures anywhere on the path.
  This will be applied in this section to the addition of \chemical{CCl_2}
  to ethylene:
  \hbox{
  \chemical{CCl_2} + 
   \chemical{H_2C,DOUBLE,CH_2}
  \chemical{GIVES}
  \startchemical[frame=off,height=fit,width=fit]
  \chemical[THREE,B2,-SB1,+SB3,Z1][CCl_2]
  \stopchemical}.
 We will first optimize the structure of the transition state and the product at 300K,
  and compare the geometrical parameters to those of the 0K geometries. 
  We will then focus on constructing the minimum free energy path at 0K and 300K.
  
 This reaction has already theoretically 
been studied in our group,\cite{KSZ04} 
as well as in  other groups.\cite{RHM80,KYM99} In particular, 
possible deficiencies of the DFT methods to describe the long range
interactions have already been stressed. However, our goal here
is to construct the reaction path on the free energy surface 
for this reaction and to compare it with the path obtained
with a predefined reaction coordinate.
 In such a comparison, the accuracy of DFT is not the main issue.

Previous studies have shown that the reaction proceeds in two steps:
 the first phase corresponds to the electrophilic addition
of the carbene to \emph{one} of the carbons making up the double bond. This phase
proceeds through a transition state that has been optimized.
The final phase is a nucleophilic attack on the second carbon of the
double bond to close the cycle. This phase proceeds without any barrier, directly
after the first one:

\begin{center}
\includegraphics{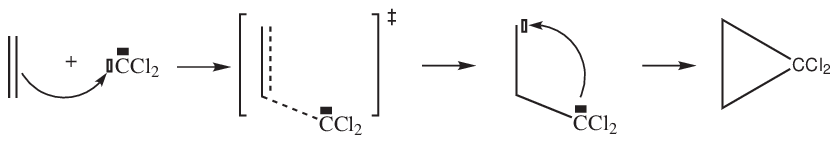}
\end{center}

As already noticed,\cite{KSZ04} the distance between the center
 of the double bond and the carbon of the carbene is a good reaction
coordinate for the first phase, but is not sufficient to
accurately describe the second phase. Therefore, we have 
 decided to include three coordinates into the active set:
$d_{CC}$: the ethylene CC bond distance,
 $d_{CG}$: the distance between the carbon atom of the carbene and
the center of the double bond denoted by $G$, and $\alpha$ the angle between the
 double bond and this last distance. This is depicted on the following scheme:

\begin{center}
\includegraphics{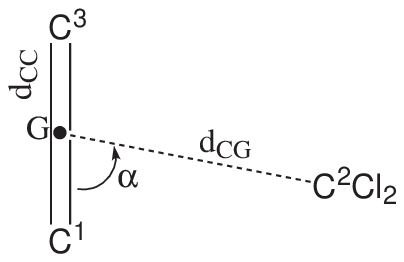}
\end{center}

For further reference,  the previous  study
using only the $d_{CG}$ distance as a reaction coordinate
 will be referred
to as the 1D study by opposition to the present study
that uses a 3 coordinates active set and will thus be referred
to as the 3D scheme.

The $\vec{Z_{\xi}}$ matrix corresponding to the constraints
 $d_{CG}$, $d_{CC}$ and $\alpha$  reads:
\begin{equation}
\vec{Z_{\xi}} =
\begin{pmatrix}
 Z_{d_{CG}} & 0 & 0\\
 0 & Z_{d_{CC}} & 0\\
0 & 0 & Z_{\alpha}  \\
\end{pmatrix}
\end{equation}

This form the the $\vec{Z_{\xi}}$ matrix is quite different from that obtained in
eq. \eqref{eqZddangle} for a system of two bonds sharing a common atom.
This comes from the fact that the elements of this matrix are defined
with respect to the coordinates of the atoms \chemical{C^1}, 
 \chemical{C^2} and  \chemical{C^3} whereas  $\alpha$
and $d_{GC}$ are defined with respect to \chemical{C^1}, 
 \chemical{C^3} and  \chemical{G}.
As  \chemical{G} is the center of mass of  \chemical{C^1C^3}, it plays
a symmetric  role in the expressions of the elements of the 
$\vec{Z_{\xi}}$ matrix, leading  to compensating terms which
sum up to zero. As an example, let us consider the off diagonal term
$Z_{d_{CC}d_{CG}}$ 
between the double bond  \chemical{C^1C^3} and the distance
\chemical{GC^2}. Simple algebra gives
\begin{align}
\frac{\partial d_{CC}}{\partial \vec{x_1}} & = - \frac{\partial
  d_{CC}}{\partial \vec{x_3}}  \\
\frac{\partial d_{CC}}{\partial \vec{x_2}} & = 0 \\
\frac{1}{m_1}\frac{\partial d_{CG}}{\partial \vec{x_1}} & =
\frac{1}{m_3}\frac{\partial d_{CG}}{\partial \vec{x_3}} 
\end{align}
which leads to 
\begin{equation}
Z_{d_{CC}d_{CG}} = \sum_{i=1}^{i=3} \frac{1}{m_i} \frac{\partial
  d_{CC}}{\partial \vec{x_i}}  \frac{\partial d_{CG}}{\partial
  \vec{x_i}}
 =   \frac{\partial   d_{CC}}{\partial \vec{x_1}} \left( 
\frac{1}{m_1}\frac{\partial d_{CG}}{\partial \vec{x_1}}
- \frac{1}{m_3}\frac{\partial d_{CC}}{\partial \vec{x_3}} \right)
= 0
\end{equation}

Similar cancellations appear for the other non diagonal terms.

We will now give the expression of $Z_{d_{CG}}$, $Z_{d_{CC}}$
 and $Z_{\alpha}$. Even though in our case all three atoms have the same mass,
 we will write the following
formula for the general case where all three atoms have different masses and
G is the center of mass of \chemical{C^1C^3}. Tedious but straightforward algebra leads to:
\begin{align}
\vec{x}_{G} &=\frac{m_1 \vec{x}_1 + m_3 \vec{x}_3}{m_1 +m_3}\\
Z_{d_{CG}} &= \frac{1}{m_2} + \frac{1}{m_1 + m_3} \\
Z_{d_{CC}} &= \frac{1}{m_1} + \frac{1}{m_3} \\
Z_{\alpha} &= \frac{Z_{d_{CG}}}{d_{CG}^2}+\frac{Z_{d_{CC}}}{d_{CC}^2} \\
\det{\vec{Z_{\xi}}} &= Z_{\alpha} Z_{d_{GC}} Z_{d_{CC}} 
\end{align}

Finally, derivatives of the free energy read:
\begin{align}
\frac{\partial A}{\partial d_{CG}} &= \frac{1}{Q}
\left( \left< \lambda_{d_{CG}} \right>
- \frac{k <T>}{ Z_{\alpha}}\frac{Z_{dGC}}{d_{CG}^3} \right)\label{eq:DerACX1} \\
\frac{\partial A}{\partial d_{CC}} &= \frac{1}{Q}
\left( \left< \lambda_{d_{CC}} \right>
- \frac{k <T>}{ Z_{\alpha}}\frac{Z_{dCC}}{d_{CC}^3} \right) \label{eq:DerACX2}\\
\frac{\partial A}{\partial \alpha} &= \frac{1}{Q} 
 \left< \lambda_{d_{CG}} \right>  \label{eq:DerACX3}
\end{align}

\subsection{Stationary points}
 \subsubsection{Transition State}
 Optimization of the transition state structure was carried out by employing
 the quasi-Newton scheme\cite{SchlegelNR}
 using the formula proposed by  Bofill to update the
Hessian.\cite{Bofill94}
 This procedure will be described in details elsewhere.\cite{YHFZ04} 
This quasi-Newton scheme is an iterative procedure that requires
 initial  values of the three parameters
$d_{CG}$, $d_{CC}$ and $\alpha$. These values were taken from the previous
1D study,\cite{KSZ04}  in which
the transition state (TS) was located at
 $d_{CG} = 4.5$ $a_0$: this  was taken as the initial
 value of $d_{CG}$ for
 our 3D optimization.  The thermal average of  the ethylene bond length
and of the angle during the previous 1D  simulation with $d_{CG} =
 4.5$ $a_0$ were taken as  initial  values for $d_{CC}$ and for the
angle $\alpha$.  The Hessian at this initial geometry 
was calculated employing finite differences of the gradients.
Diagonalization lead to only one negative eigenvalue proving
 the transition state nature of the starting point.
The Hessian matrix was also calculated and diagonalized for the final
geometry: we found only one negative eigenvalue, corresponding to
an eigenvector directed mainly along the $d_{CG}$ variable.

\begin{figure}[!ht]
\includegraphics{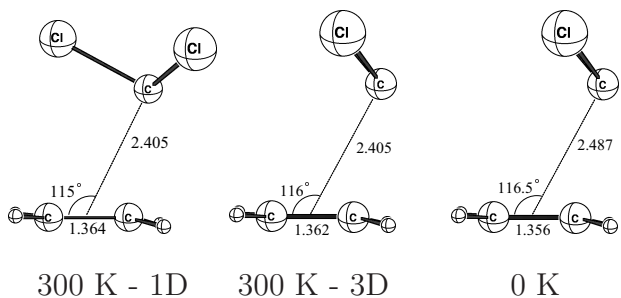}
\caption{Transition states geometries for the addition
of dichlorocarbene {\protect \chemical{CCl_2}} to ethylene,
 optimized at 0K,
 at 300K with only the $d_{CG}$ distance constrained (1D),
and at 300K with $d_{CG}$, $d_{CC}$ and $\alpha$ all  constrained
(3D). $G$ is the midpoint of the double bond.  Distances in \AA, angles  in deg.} \label{Fig:TS}
\end{figure}

 The resulting geometry is reported on Figure \ref{Fig:TS}, together
 with the geometry obtained at 300K when only $d_{CG}$ is constrained,
 as well as the geometry obtained on the PES at 0K. The main geometrical
 parameters are given in Table \ref{Tab:TS}.

\begin{table}[!ht]
\begin{tabular}{c|ccc}
 & $d_{CG}$ (\AA) & $d_{CC}$ (\AA) & $\alpha$ (deg.)\\
\hline
1D &  2.405(2)$^{(a)}$ & 1.364(2)$^{(b)}$ & 115(1)$^{(b)}$ \\
3D & 2.405(2) & 1.362(2) & 116(1) \\
\hline
0K & 2.487 & 1.356 & 116.5 \\
\end{tabular}
\caption{Main geometrical parameters for the transition state
geometries for the addition
of the dichlorocarbene {\protect \chemical{CCl_2}} on ethylene, 
 optimized at 0K,
 at 300K with one constraint (1D),
and at 300K with three constraints (3D). (a) Estimate of the
uncertainty. The uncertainty for structures on the potential energy
surface is much smaller, and thus not quoted.
 (b) Average values. } \label{Tab:TS}
\end{table}

 All three structures are non symmetric: the $\alpha$ angle
is approximately equal to 115\degres{} instead of 90\degres.
This is in agreement with the Woodward-Hoffman rules and with 
previous calculations\cite{HRM84,BWJ89,KYM99,BCK04,KMSH99} and experimental
results.\cite{KMSH99} More, all geometries correspond to an early
transition state in which the ethylene molecule 
is only slightly distorted. The double bond length equals
approximately 1.36 \AA, close to the equilibrium distance in
the free molecule: $d_{CC} = 1.33$ \AA.
 
The two structures found at 300K are almost identical.
 This is not surprising: 
 as already stated the distance $d_{CG}$ between the middle of
the double bond and the carbon atom of the carbene is a good reaction
coordinate up to this point.

The transition state geometry at 0K is in fairly good agreement with
previous \emph{ab initio} calculations\cite{HRM84,BWJ89,KYM99,BCK04}
 except for the $d_{CG}$ distance
which is slightly overestimated here: $d_{CG} = 2.487$~\AA{} compared 
to $d_{CG} = 2.37$~\AA{} at the B3LYP/6-31G* level,\cite{BCK04}
 or $d_{CG} = 2.38$~\AA{} at the MP2/6-31G* level.\cite{BWJ89,KYM99,BCK04} 
 This is due to the fact that this 
region of the potential energy surface is shallow so that the location
of the transition state depends strongly on the functional and on
the basis set used. 
 Similarly, the $\alpha$ angle
is also a bit overestimated: it equals 116.5\degres{} in our study,
whereas it equals respectively 111.7\degres{} and 112.2\degres{} at
the MP2/6-31G* and B3LYP/6-31G* levels of calculation.\cite{BWJ89,KYM99,BCK04}

Going from the transition state structure found at 0K to that obtained
at 300K leads to
an increase of the double bond length due to thermal vibrations.
On the other hand, the $d_{CG}$ distance is smaller at hight temperature.
This comes from the fact that the barrier originates mainly
from rotational entropy lost when the transition state is formed.
As the average rotational momentum increases with the temperature,
we expect this barrier to move to smaller distance as the temperature
rises. This is in agreement with previous studies.\cite{KSZ04,YHFZ04}

\subsubsection{1,1'-dichlorocyclopropane}

\begin{figure}[!ht]
\includegraphics{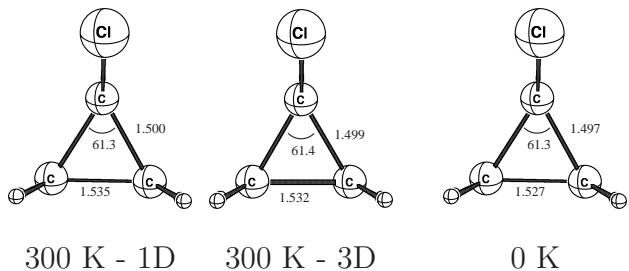}
\caption{Geometries of the 
1,1'-dichloro cyclopropane 
 optimized at 0K,
 at 300K with only the $d_{CG}$ distance constrained (1D),
and at 300K with $d_{CG}$, $d_{CC}$ and $\alpha$ all three constrained
(3D). $G$ is the midpoint of the double bond.   Distances in \AA, angles in deg.} \label{Fig:Min}
\end{figure}

The main geometrical parameters for the 1,1'-dichlorocyclopropane 
 are reported in Table \ref{Tab:Min}
and the corresponding structures are given in Figure \ref{Fig:Min}.
All three structures are very symmetric: $\alpha = 90$\degres{}
which means that the carbon atoms form an isocel triangle,
 its base being the former ethylenic bond. All
three carbon-carbon bond length are now close to that of
a single bond:
 $d_{C^1C^3} = 1.53$ \AA, $d_{C^1C^2} = d_{C^3C^2} = 1.50$
 \AA.

\begin{table}[!ht]
\begin{tabular}{c|ccc}
 & $d_{CG}$ (\AA) & $d_{CC}$ (\AA) & $\alpha$ (deg.)\\
\hline
1D &  1.289(2)$^{(b)}$ & 1.535(2)$^{(b)}$ & 90(1)$^{(b)}$ \\
3D & 1.289(2) & 1.532(2) & 90(1) \\
\hline 
 0K &   1.288 & 1.527 &  90.1 \\
\end{tabular}
\caption{Main geometrical parameters for the 1,1'-dichlorocyclopropane,
 optimized at 0K, at 300K with one constraint (1D),
and with three constraints (3D). (a) Estimate of the
uncertainty.  (b) Average values.} \label{Tab:Min}
\end{table}

Similarly to what was observed for the transition state structures, there
are very little difference between the room temperature 1D and 3D structures. This
comes from the fact that the gradient is zero for a minimum, so that
the definition of the reaction coordinate does not matter anymore.
To further assess this point, we have launched a simulation with no constraints.
The average values of $d_{CG}$, $d_{CC}$ and $\alpha$ are in very good
agreement with the constrained simulations: $d_{CG}=1.291 \pm 0.002$ \AA,
$d_{CC} = 1.533 \pm 0.002$ \AA{} and $\alpha=90. \pm 1$\degres.

The 0K structure is in  good agreement with previous calculations:
the difference in  distances and angles is less then 0.01 \AA, and
1\degres, respectively.\cite{HRM84}
When going from 0K to 300K,  the C-C bonds elongate, while 
$d_{CG}$ and $\alpha$ remain constant. This comes from the fact that
this molecule has a \chemical{C_{2v}} symmetry which imposes
that the average angle should be close to
its value on the potential energy surface.

\subsection{Reaction Path}

We focus now on the core of this application: the construction
of the reaction path  connecting the reactants to the product, directly on the free energy surface.
The aim of this part is dual: first, we show how to use the previous
formula on a real example. Second, we compare the 3D path constructed
here to the path obtained with only one `chemically intuitive' reaction coordinate.
   Starting from the transition state, we have constructed the
 forward reaction path   leading to the product and the backward path leading
  to the reactants.

 In this work,   we have  moved along the gradient employing a small step size:
 at each point $\vec{x^k}$ of the path, a simulation is launch while constraining all three active
 coordinates $d_{CG}$, $d_{CC}$ and $\alpha$. The previous formulas (\eqref{eq:DerACX1},
 \eqref{eq:DerACX2} and \eqref{eq:DerACX3}) are used to compute the
 free energy gradient: $\vec{g} = \left( \frac{\partial A}{\partial d_{CG}}, 
 \frac{\partial A}{\partial d_{CC}}, \frac{\partial A}{\partial \alpha} \right)^t$.
 We then convert this gradient into a normalized mass weighted gradient:
 $\vec{g}_{MW} = \mathcal{N} \vec{Z_{\xi}} \vec{g}$, 
 with $\mathcal{N}=(\vec{g}^t \vec{Z_{\xi}} \vec{g})^{\sfrac{-1}{2}}$. The next point $\vec{x^{k+1}}$ is
 calculated by following the gradient on a small distance $ds$:
 \begin{equation}
 \vec{x^{k+1}} =  \vec{x^{k}} - ds \times \vec{g}_{MW}
 \end{equation}
    An alternative way is to employ the scheme derived by Gonzalez \etal\cite{GS89}
  which allows for the use of a much bigger stepsize.
The forward path corresponds to the closure of the cycle, that is to say
to the formation of the second carbon-carbon bond. The gradient
along this path is large, and we used a stepsize of 0.5 
a.u.\cite{footnote1}
The backward path corresponds to the departure of the carbene, which
is the reverse of the electrophilic addition. The change in free energy is small in 
that direction, and we had to employ a smaller stepsize of 0.2 a.u. to
minimize the statistical noise.

\subsubsection{Free energy profile}

\begin{figure}[!ht]
\includegraphics{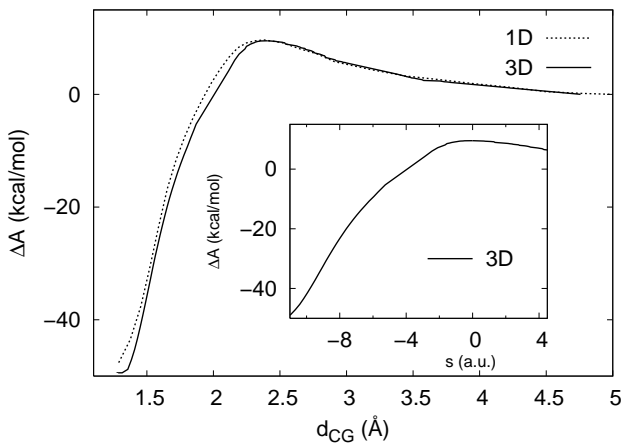}
\caption{Free Energy Profile (FEP) for the  addition
of dichlorocarbene {\protect \chemical{CCl_2}} to ethylene.
The \textbf{1D} profile is calculated using $d_{CG}$ as the 
predefined reaction coordinate, whereas the \textbf{3D} profile
uses an active set of three coordinates  ($d_{CG}$, $d_{CC}$,
$\alpha$).  The inset shows the FEP plotted against the
curvilinear distance from the transition state along the 3D path.
$G$ is the midpoint of the double bond.} \label{Fig:FEP}
\end{figure}

The resulting free energy profile (FEP) is reported in Figure \ref{Fig:FEP}
together with the profile generated with one constraint.\cite{KSZ04} In order to
compare them, both paths have been plotted using the $d_{CG}$ distance
as an approximate reaction coordinate. The FEP obtained in the present
work is given as an inset in Figure \ref{Fig:FEP}.
 The two profiles are very similar: first, the free energy increases smoothly
 from the isolated reactants to the transition state. Then, it
 decreases abruptly when the cyclopropane is formed. 

On an energetic ground, 
in agreement with the fact that the  free energy is a state function, 
both path lead to a similar change in free energy equal to:
 $\Delta A_{corr}$ = -47.6 kcal$\cdot$mol$^{-1}$ in this work, and
 $\Delta A_{corr}$ = -46.5 kcal$\cdot$mol$^{-1}$ in our previous work, when employing 40000 steps.
 More, as $d_{CG}$
is a good reaction coordinate for the first phase, the barrier
is also similar in both cases: 
 $\Delta A^{TS}_{corr} \approx$ 11.5 kcal$\cdot$mol$^{-1}$ in this work, compared to
$\Delta A^{TS}_{corr} \approx$ 11.7 kcal$\cdot$mol$^{-1}$
previously. These values are in good agreement with the previous
studies.\cite{KSZ04,BWJ89}

\subsubsection{Geometrical parameters}

\begin{figure}[!ht]
\includegraphics{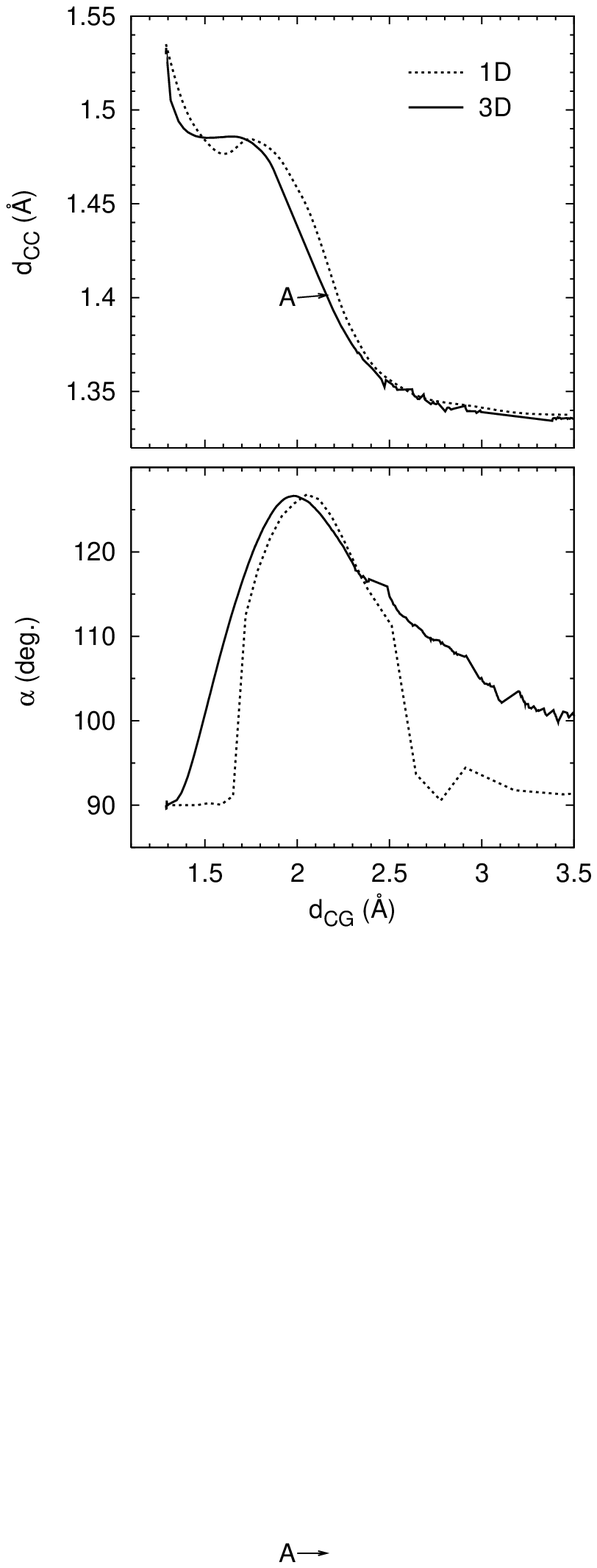}
\caption{Reaction path for the  addition
of dichlorocarbene to ethylene. Two projections
are show: in the $d_{CC}-d_{CG}$ and $\alpha-d_{CG}$
subspace, $G$ being the midpoint of the double bond.
The \textbf{1D} path is obtained  using $d_{CG}$ as the
reaction coordinate, whereas the \textbf{3D} path
uses an active set of three coordinates ($d_{CG}$, $d_{CC}$,
$\alpha$).} \label{Fig:Path}
\end{figure}

Even thought the two free energy profiles are similar,
 the two path are actually quite
different in terms of geometrical parameters. 
We discuss here the variations of the structural parameters
$d_{CC}$ and $\alpha$ for both path.
 The evolution of the double bond distance $d_{CC}$ and of the $\alpha$
angle obtained in this study are reported on Figure  \ref{Fig:Path}
 together 
with the average values $\left< d_{CC} \right>$ and $\left< \alpha
 \right>$
obtained in the previous study using only $d_{CG}$ as a reaction
 coordinate.

The first feature worth noting is that  the forward path,
 corresponding
to the  the formation of the 
second C-C bond,  is not parallel to the $d_{CG}$ axis
but acquires contributions along both $d_{CC}$ and $\alpha$. This confirms that
an accurate description of this step requires a more complex reaction
coordinate than just the $d_{CG}$ distance.
As a consequence, despite the fact 
that $d_{CG}$ varies only from 2.40 \AA{} to 1.29 \AA{} for the
second phase, the actual path length 
is of the same order of magnitude
as for the first phase: $ s \approx 10$ a.u.

The evolution of the two structural parameters $d_{CC}$ and $\alpha$
can be divided into three zones. The first one corresponds to $d_{CG}
\gtrsim 2.6$ \AA. In this zone, there is little interaction between
the two reactants: $d_{CC}$ is constant and equal to 1.33 \AA{}, which
is the standard double bond length, and $\alpha$ tends to 
90\degres.\cite{footnote2}
This comes from the fact that, for the unconstrained system, at large
separation there is free rotation of the carbene  around
the ethylene molecule. As a consequence, $\alpha$ should vary freely
between 0\degres{} and 180\degres, with an average value of 90\degres.

The second zone corresponds to the electrophilic addition, that is the
formation of the first bond between the carbene and the ethylene
molecule. In this zone, $d_{CG}$ evolves from ca. 1.8 \AA{} to ca.
2.6 \AA{}.
As the two molecules start to interact the ethylenic bond elongates
from 1.33 \AA{} to ca. 1.48 \AA. As expected, this last value is
intermediate between a single and a double carbon-carbon bond length.
Simultaneously, $\alpha$ increases to 128\degres. This is a
consequence of the fact that the symmetric approach for which
$\alpha=90$\degres{}
is forbidden, and is thus associated with a very high
barrier.\cite{KMSH99}
During this phase, the $C^1C^2$ bond is formed: this distance decreases from 2.4 \AA{} 
to 1.56 \AA, which is close to a single bond.

The last zone corresponds to $d_{CG} \leq 1.8$ \AA{} and represents
 the closure of the cyclopropane ring. The $C^1C^3$ bond (formerly
the double bond) length increases to its finals value of 1.538 \AA,
while $\alpha$ drops to 90\degres. Comparing the path constructed
using a 3 coordinates active space to the path previously obtained
shows that $\alpha$ is a much better reaction coordinate for this
phase than $d_{CG}$. Indeed, during this phase the $C^1C^2$ distance
is almost constant and close to 1.54 \AA{}, while the $C^3C^2$ distance
decreases strongly from 2.3 \AA{} to 1.54 \AA. This illustrates that
the second phase of the reaction is actually the bending of
the $C^1C^2$ bond towards the $C^3$ carbon atom. For this type of
movement $\alpha$ is a  good reaction coordinate whereas $d_{CG}$ is
a poor one. As a consequence, when only $d_{CG}$ is used as a reaction
coordinate, $\alpha$ drops abruptly from 115\degres{} to 90\degres{}
around  $d_{CG} = 1.36$ \AA. On the other hand, the variations
of  $\alpha$ are much smoother with our active set.

To conclude this section, we would like to point out that, as expected,
 both path are qualitatively similar, with
the same overall evolutions of the geometrical parameters.
However, they differ significantly on a quantitative basis, especially
 in the third zone. This is due to the fact that $d_{CG}$ is not a good approximation to the
reaction coordinate in this zone.
To illustrate how this explains why the two path are different, let us
 consider a force acting on the $d_{CC}$ variable at the point A (Fig. \ref{Fig:Path}) 
located in this zone. We denote the local
coordinate set by $(\xi, u, v)$, with $\xi$ being the reaction coordinate.
From the definition of a reaction coordinate, we have:
\begin{gather}
\frac{\partial A}{\partial \xi} = g \not= 0 \\
\frac{\partial A}{\partial u} =  0 \\
\frac{\partial A}{\partial v} =  0 
\end{gather}

As $d_{CG}$ is not a good approximation to the reaction coordinate,
$\xi$ depends not only on $d_{CG}$ but also on $d_{CC}$ and $\alpha$.
As a result, the derivative of the free energy along $d_{CC}$ is
related to the gradient of the free energy:
\begin{align}
\frac{\partial A}{\partial d_{CC}} & = 
\frac{\partial A}{\partial \xi}  \frac{\partial \xi}{\partial d_{CC}}
+ \frac{\partial A}{\partial u}  \frac{\partial u}{\partial d_{CC}}
+ \frac{\partial A}{\partial v}  \frac{\partial v}{\partial d_{CC}} \notag \\
& = \frac{\partial A}{\partial \xi}  \frac{\partial \xi}{\partial d_{CC}} \not= 0 
\end{align}

Thus, if $d_{CC}$ is not constrained, the system will evolve in order
to minimize the free energy in that direction, and the two paths
 will not coincide. 

\subsection{Importance of the correction to evaluate the gradients}

\begin{figure}[!ht]
\includegraphics{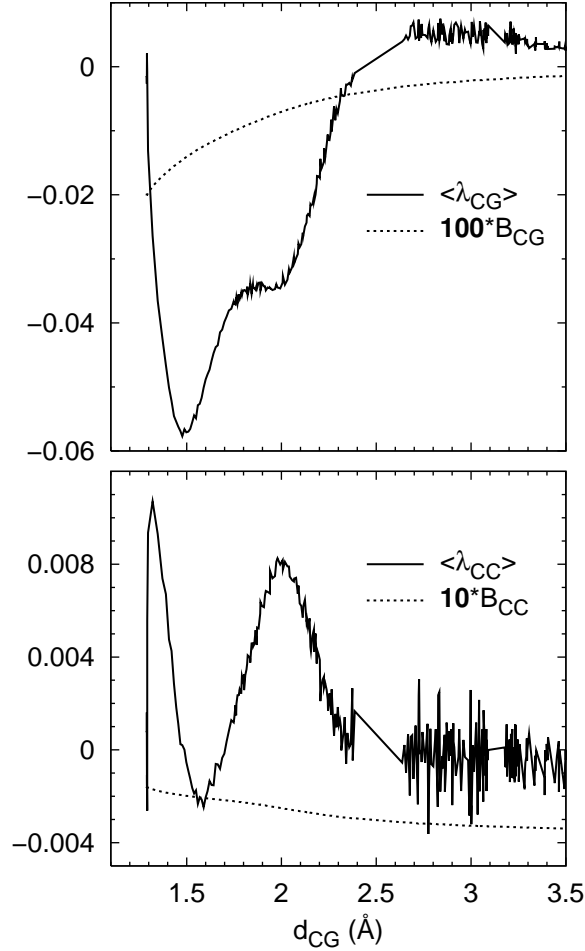}
\caption{Importance of the correction terms in the evaluation
of the free energy derivative. $\left< \lambda_X
  \right>$ is the average of the Lagrange multiplier. 
$B_{CG} = -kT\frac{Z_{d_{CG}}}{Z_{\alpha}d^3_{CG}}$  and
$B_{CC} = -kT\frac{Z_{d_{CC}}}{Z_{\alpha}d^3_{CC}}$ are the correction
to the derivative of the free energy along $d_{CG}$ and $d_{CC}$
respectively.  $G$ is the midpoint of the double bond. 
} \label{Fig:Bterms}
\end{figure}

It is worth analyzing the importance of the B terms
in equations \eqref{eq:DerACX1}-\eqref{eq:DerACX3}.
 They are reported on Figure \ref{Fig:Bterms} together with the
average value of the Lagrange multipliers. It appears that those terms are
usually smaller than the Lagrange multiplier by at least an order of magnitude.
However, as the reaction path is following the free energy gradient, these
small differences are accumulated along the path, leading to a non-negligible correction.
For example, for the studied reaction, the free energy correction along the path can reach 
 1 kcal$\cdot$mol$^{-1}$.
 The larger correction for the geometrical parameters is
 observed for the evolution of the double bond length:
 $\Delta(d_{CC}) \approx 0.01$ \AA.
However, the structural differences for  the product and the transition
state are much smaller: $\Delta(d) \leq 0.005$ \AA{} and $\Delta(\alpha) \leq 0.5$\degres.
The correction for the total free energy difference
is only $\Delta(\Delta A_{corr})=0.1$ kcal$\cdot$mol$^{-1}$.

We would like to draw attention to the fact that, even thought those terms
are generally not negligible, one should not forget that we are using
classical mechanics, and that all quantum effects are missing for
 the description of the nuclei motion. In particular, ZPE and tunneling
are completely neglected.

\subsection{Effect of the temperature}

\begin{figure}[!ht]
\includegraphics{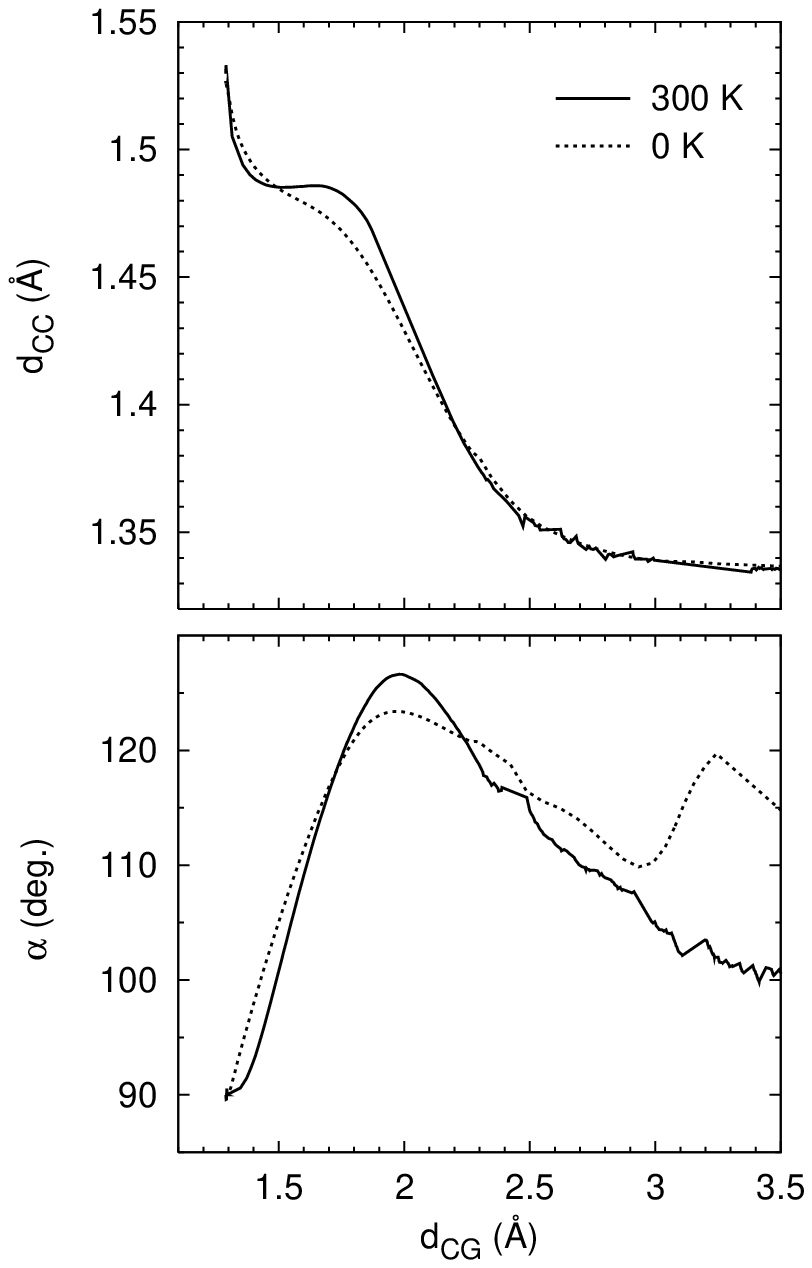}
\caption{Reaction path for the addition of the dichlorocarbene to ethylene at
0K and 300K. Two projections are show: in the $d_{CC}-d_{CG}$ and $\alpha-d_{CG}$
subspace, $G$ being the midpoint of the double bond.} \label{Fig:Path0}
\end{figure}

Following the procedure of Gonzalez \etal,\cite{GS89} we have constructed
the reaction path at 0K. 
As the differences between
 the potential energy profile and the free energy profile have already
been discussed,\cite{KSZ04} we shall focus here on the differences between
the path at 0K and 300K. The two path are reported on figure \ref{Fig:Path0}.

At large separation, that is for the electrophilic
 addition, 
 $d_{CG}$ is a good reaction coordinate  at both temperatures. More, as long as
 the two reactants interact only weakly, the thermal motions are mainly vibrations
 of the two molecules. As the potential energy surface is almost harmonic for
 such vibrations, the  evolutions
of the double bond distance and of the angle $\alpha$ are 
similar at 300K and at 0K. On the potential energy surface,
as the distance between the two fragments increases, the interaction
depends less and less on the angle, leading to some spurious evolutions.
On the free energy surface, this independence of the interaction
on the angle leads to an average value of 90\degres{} instead.

When the interaction starts to be stronger, these oscillations of the
angle disappear and both path are qualitatively similar.
However, due to the vibrational thermal energy, at 300K the 
parameters changes are larger.

\subsection{Computational details}

The Car-Parrinello projector augmented-wave (CP-PAW)\cite{PAW1,PAW2}
program by Bl{\"o}chl was used for all AIMD calculations. 
In the CP-PAW calculations, periodic boundary conditions
were used in all examples with an orthorhombic unit cell
described by the lattice vectors ([0, 14.74, 14.74], [14.74, 0,
14.74], [14.74, 14.74, 0]) (bohr, 7.8 \AA).
 The energy cutoff used to define the basis
set was 30 Ry (15 a.u.) in all cases. Because the systems of
interest are all isolated molecules, only the $\Gamma$-point in k-space
was included and the interaction between images was removed
by the method proposed by Bl\"ochl.\cite{PAW2}
 The approximate density-functional
theory (DFT) used here consisted of the combination
 of the Perdew-Wang parametrization of the electron gas\cite{PW91}
in combination with the exchange gradient correction presented by Becke\cite{B88}
 and the correlation correction of Perdew.\cite{PExc}
The SHAKE algorithm\cite{SHAKE}
 was employed in order to impose the constraints. The
mass of the hydrogen atoms was taken to be that of deuterium, and
normal masses were taken for all other elements. 

Room temperature CP-PAW calculations were performed at a target
temperature of 300 K.
The Andersen thermostat\cite{Andersen80} was applied to the nuclear motion
by reassigning the velocity of N randomly chosen nuclei every
n steps where N and n are chosen to maintain the desired
temperature. In our case this
amounted to one velocity reassignment every 20 steps.
 Thermostat settings were monitored and adjusted if necessary during
 the equilibration stage, with the main criteria for adequate thermostating
 being the mean temperature lying within a range of $300 \pm 10$ K and a 
temperature drift lower than 1 K/ps. In combination with the 
Andersen thermostat, a constant friction was applied to the wave
function with a value of 0.001. 
Following the conclusions of the previous study, for each simulation,
we performed between 35000 and 50000 steps in order to ensure that the system 
was fully equilibrated and that the temperature and the free energy
gradient were fully converged.

The free energy profiles were obtained by numerical integration
of the gradient along the path, using a procedure similar 
to the pointwise thermodynamic integration (PTI) method.\cite{SC91}
As the overall rotation and translation of the molecule are frozen
during a simulation, one has to correct the free energy obtained
from a simulation. We have used the procedure of Kelly \etal\cite{KSZ04}
To summarize, the overall 
correction for the entropy is the sum of the translational
and rotational entropy:
\begin{equation}
 \Delta S^{AB}_{corr}(s)=S_R^{AB}(s)+S_T^{AB}(s) - S_T^{A}(\infty)
-S_T^{B}(\infty) 
\end{equation}
Where $S_R^{AB}(s)$  is the rotational entropy at RC = s
 which is geometry dependent, and  $S_T^{AB}(s)$ is
 the translational entropy at RC =s. The last two
 terms represent the translational entropy of the isolated species
 A and B.
These terms are calculated using standard formula for the partition
functions.
Finally,  the total free energy change  $ \Delta A^{AB}_{corr}(s)$
is obtained from a CP-PAW
 simulation with the constraints described above as 
 $$ \Delta A^{AB}_{CM}(s) = \Delta A^{AB}_{PAW}(s) - T  \Delta S^{AB}_{corr}(s)$$
where $\Delta A^{AB}_{PAW}(s)$ is the change in free
 energy obtained directly from the simulation,
and CM (classical mechanics) 
refers to the fact that the motion of the nuclei is described
 using classical mechanics.
It should be mentioned that the zero-point energy (ZPE)
 correction is not included in our simulations.  This should
 not seriously hamper our objective which is to analyze
 the qualitative differences between the path obtained with one and
three constraints.

\section{Conclusion} \label{SecConcl}

In this work, we have proposed a new look at the standard formulas
for evaluating the derivatives of the free energy along a reaction
coordinate.

First, we recollected the different formulas available in a uniform
approach. These formulas allow one to compute accurately and efficiently
the gradient of the free energy for an \emph{unconstrained} system
using a \emph{constrained} molecular dynamic simulation.

The main finding of this investigation is a set of equations that makes it
possible to construct a minimum free energy reaction path
 instead of calculating
the free energy changes along a predefined path.
Indeed, we believe that one can use the free energy gradients in the same
way potential energy gradients have been used in the past 20 years in quantum chemistry
calculations.

Addition of the dichlorocarbene to ethylene was studied as a numerical
example. It was shown that a simulation using only one
constraint is not sufficient to describe the whole path. Using
the free energy gradient in a subset of three active coordinates 
lead to a smoother path, refining the understanding of this process.

\begin{acknowledgments}
The authors would like to thank Dr. Michael Seth and Dr. Shengyong Yang for helpful
and fruitful discussions. This work was supported by the National Sciences and Engineering
 Research Council of Canada (NSERC). Calculations were performed in part on 
 the Westgrid cluster and the MACI Alpha cluster located at the University of Calgary.
One of us (TZ) thanks the Canadian government for a Canada Research Chair. 
\end{acknowledgments}

\appendix
\section{Properties of the Mass matrix} \label{AppMassMat}
  Using $\vec{A}_{q\xi}^{-1} \vec{A}_{q\xi} = \vec{I}$, one readily finds:
\begin{align}
 \vec{X}_{q}\vec{A}_{q}+ \vec{Y}_{\xi}\vec{B_\xi}^t & = \vec{I_{3N-1}} \label{Aq1} \\
 \vec{X}_{q}\vec{B_\xi} + \vec{Y_\xi}C_{\xi} & = \vec{0} \label{Aq2} \\
\vec{Y}_{\xi}^t\vec{A}_{q}+ Z_\xi \vec{B}_{\xi}^t & = \vec{0} \label{Aq3} \\
 \vec{Y}_{\xi}^t\vec{B_\xi} + Z_\xi C_{\xi} & = 1 \label{Aq4}
\end{align}

Plugging equation (\ref{Aq3}) into equation (\ref{Aq1}) leads to
\begin{equation}
\vec{A}_{q}^{-1} =\vec{X}_{q}-
\vec{Y_\xi}Z_{\xi}^{-1}\vec{Y}_{\xi}^t \label{eqAqu}
\end{equation}

 The last relation we need derives from:
\begin{equation}
\left(
 \begin{array}{cc}
 \vec{A}_{q} & \vec{0} \\
\vec{B}_{\xi}^t & 1 \\
\end{array} \right)
= \vec{A}_{q\xi} \vec{A}_{q\xi}^{-1} \left( \begin{array}{cc}
 \vec{A}_{q} & \vec{0} \\
\vec{B}_{\xi}^t & 1 \\
\end{array} \right)
=\vec{A}_{q\xi} \left(
\begin{array}{cc}
 \vec{I}_{3N-1} & \vec{Y}_{\xi} \\
\vec{0} & Z_{\xi} \\
\end{array} \right)
\end{equation}

 Equating the determinant of the first and last term, we get:
\begin{equation}
\det{\vec{A}_{q\xi}} = \det{\vec{A}_{q}}\det{Z_{\xi}}^{-1}
= \det{\vec{J}^t \vec{M} \vec{J}} = \det{\vec{J}}^2\det{\vec{M}}
\label{eqDetAqZ}
\end{equation}

\section{Properties of Gaussian integrals} \label{AppGint}
We recall here the main properties of Gaussian integrals:
\begin{align}
 \int \vec{du} e^{\left(-\vec{u}^t \vec{A} \vec{u}\right)}
& \propto \det{\vec{A}}^{-\sfrac{1}{2}} \label{eq:Gint0} \\
 \int \vec{du} e^{\left(-\vec{u}^t \vec{A} \vec{u}\right)}
\vec{u}^t  \vec{B} \vec{u} & \propto \frac{1}{2}
 \text{Tr}\left(\vec{A}^{-1} \vec{B}\right)
 \int \vec{du} e^{\left(-\vec{u}^t \vec{A} \vec{u}\right)}  \label{eq:Gint2}
\end{align}

\section{Constraining the overall rotation and translation} \label{AppTR}
 In this section, we derive the expressions used to constrain the overall translation and rotation of
the system. For this, we start from a reference geometry $\vec{x^0}$ and we seek the conditions that
should be satisfied by the new geometry $\vec{x}$.
 The Center of Mass G is defined by:
 
 $$ \vec{x_{G}}=\frac{\sum_{i=1}^{N}m_{i}\vec{x_{i}}}{\sum_{i=1}^{N}m_{i}}$$
 where $m_i$ is the mass of the atom $i$ with coordinate $\vec{x_i}$.
 In the following, the notation $\widetilde{\vec{x}}_{i}=\vec{x_{i}}-\vec{x_{G}}$ will be used. 
 Let us denote by  $\hat{x}$, $\hat{y}$  and $\hat{z}$ the unit vectors of  our laboratory coordinate system.
 For the sake of clarity, the component of $x_i$ along the  $\hat{x}$, $\hat{y}$  and $\hat{z}$  axis will
be denoted by $x_{i,1}$, $x_{i,2}$ and $x_{i,3}$ respectively.

Constraining the translation is equivalent to freeze the movement of
the center of mass, i.e. to apply the following linear constraints:
\begin{subequations}
\begin{gather}
\sigma_{1}  =  \sum_{i=1}^{N}\frac{m_{i}}{M} \left(x_{i,1}-x^0_{i,1}\right) = 0 \\
\sigma_{2}  =  \sum_{i=1}^{N}\frac{m_{i}}{M} \left(x_{i,2}-x^0_{i,2}\right) = 0 \\
\sigma_{3}  =  \sum_{i=1}^{N}\frac{m_{i}}{M} \left(x_{i,3}-x^0_{i,3}\right) = 0 
\end{gather}
\end{subequations}

Constraining the rotation can be done by using the second Eckart
 conditions.\cite{Eckart35}  These conditions minimize the angular
 momentum due to small displacements: they provide an approximate way to
  constrain the global rotation during a molecular dynamic simulation:
\begin{subequations}
\begin{gather}
\sigma_{4} =  \hat{x} \cdot \left[ \sum_{i=1}^{N} m_{i} \left(\widetilde{\vec{x}^0_{i}} \times \widetilde{\vec{x_{i}}}\right) \right] = 0 \\
\sigma_{5}  =  \hat{y} \cdot \left[ \sum_{i=1}^{N} m_{i} \left(\widetilde{\vec{x}^0_{i}} \times \widetilde{\vec{x_{i}}}\right) \right] = 0 \\
\sigma_{6}  =  \hat{z} \cdot \left[ \sum_{i=1}^{N} m_{i} \left(\widetilde{\vec{x}^0_{i}} \times \widetilde{\vec{x_{i}}}\right) \right] = 0 
\end{gather}
\end{subequations}
that is: 
\begin{subequations}
\begin{gather}
\sigma_{4}  =  \sum_{i=1}^{N} m_{i} \left( \widetilde{x}^0_{i,2}\widetilde{x}_{i,3} 
- \widetilde{x}_{i,2}\widetilde{x}^0_{i,3} \right) = 0 \\
\sigma_{5}  =  \sum_{i=1}^{N} m_{i} \left( \widetilde{x}^0_{i,3}\widetilde{x}_{i,1} 
- \widetilde{x}_{i,3}\widetilde{x}^0_{i,1} \right) = 0 \\
\sigma_{6}  =  \sum_{i=1}^{N} m_{i} \left( \widetilde{x}^0_{i,1}\widetilde{x}_{i,2} 
- \widetilde{x}_{i,1}\widetilde{x}^0_{i,2} \right) = 0 
\end{gather}
\end{subequations}

 These last equations can be written as linear constraints:
 \begin{subequations}
 \begin{gather}
\sigma_{4}  =  \sum_{i=1}^{N} m_{i} \left\{ x_{i,3} 
\left( \widetilde{x}^0_{i,2} - \sum_{j=1}^{N}\frac{m_{j}}{M} \widetilde{x}^0_{j,2} \right) 
- x_{i,2} \left( \widetilde{x}^0_{i,3} - \sum_{j=1}^{N}\frac{m_{j}}{M} \widetilde{x}^0_{j,3} \right) \right\} =0  \\
\sigma_{5}  =  \sum_{i=1}^{N} m_{i} \left\{ x_{i,1} 
\left( \widetilde{x}^0_{i,3} - \sum_{j=1}^{N}\frac{m_{j}}{M} \widetilde{x}^0_{j,3} \right) 
- x_{i,3} \left( \widetilde{x}^0_{i,1} - \sum_{j=1}^{N}\frac{m_{j}}{M} \widetilde{x}^0_{j,1} \right) \right\} =0 \\
\sigma_{6}  =  \sum_{i=1}^{N} m_{i} \left\{ x_{i,2} 
\left( \widetilde{x}^0_{i,1} - \sum_{j=1}^{N}\frac{m_{j}}{M} \widetilde{x}^0_{j,1} \right) 
- x_{i,1} \left( \widetilde{x}^0_{i,2} - \sum_{j=1}^{N}\frac{m_{j}}{M} \widetilde{x}^0_{j,2} \right) \right\} =0 
\end{gather}
\end{subequations}

\section{Actually calculating eq. \ref{eqdAq12}} \label{AppCalcqn}

In this appendix, we propose one way to calculate the derivatives of the
 free energy A along  $q_n$, using a simulation  in which
 only $\xi$ is constrained.
As an example, we consider that the coordinates $q_i$ to $q_n$ 
 were inactive at step $k-1$ and become active at step $k$.
 
 The first step is to use the expressions for the generalized coordinates 
 to obtain the values for $\vec{Z_{\xi}}$, $Z_{q_n}$,
 $\frac{\partial Z_{q_n}}{\partial x'}$ and 
 $\frac{\partial q_n}{\partial x'}$.

The main difficulty in using  equation \eqref{eqdAq12} is that
one must ensure  that the sampling of $q_n$  around $q_n^k$
is sufficient. In practice, this imposes to have long  MD simulations,
with approximately 100000 steps for
each considered inactive coordinate. 
One way to circumvent this problem is to use a Taylor expansion of the derivative of the
potential around $q_i^k$, \ldots, $q_n^k$:
\begin{multline}
\frac{\partial V}{\partial q_n}
 = 
\left. \frac{\partial V}{\partial q_n^k}  \right)_{q_i^k, \ldots, q_n^k, \xi^*}
+ \sum_{j=i}^{j=n}\left. \frac{\partial^2 V}{\partial q_n \partial q_j
}  \right)_{q_i^k, \ldots, q_n^k, \xi^*} (q_j - q_j^k) \\
+ \frac{1}{2} \left. \sum_{j=i}^{j=n} \sum_{l=i}^{l=n}
\frac{\partial^3 V}{\partial q_n\partial q_j \partial q_l }
\right)_{q_i^k, \ldots, q_n^k, \xi^*} (q_j - q_j^k) (q_l - q_l^k)
\end{multline}

The simulation data is then used to fit the coefficients of this
expression. The resulting equation is then plugged into equation \eqref{eqdAq12}.


\begin{thebibliography}{10}
\bibitem{footnote1}
Even thought this value might seem quite large, because of the two
heavy chlorine atoms, a step of 0.5 a.u. corresponds to a decrease of the $d_{CG}$ distance
by approximately 0.005 \AA{} only.
\bibitem{footnote2}
The convergence is slow, and is not shown on the figure.
\end{thebibliography}
\end{document}